\documentclass[12pt]{iopart}

\usepackage{graphicx}
\usepackage{dcolumn}
\usepackage{bm}
\usepackage{amssymb}
\usepackage[colorlinks]{hyperref}
\usepackage{cite}

\newcommand{\rv}{{\bf r}}
\newcommand{\xv}{{\bf x}}

\newcommand{\Xv}{{\bf X}}
\newcommand{\Tv}{{\bf T}}
\newcommand{\Nv}{{\bf N}}
\newcommand{\Bv}{{\bf B}}
\newcommand{\ev}{{\bf e}}

\usepackage{xcolor}

\begin{document}

\title[]{Helical close-packing of anisotropic tubes}

\author{Benjamin R. Greenvall \& Gregory M. Grason}

\address{Department of Polymer Science and Engineering, University of Massachusetts, Amherst, MA 01003}
\ead{grason@mail.pse.umass.edu}
\vspace{10pt}
\begin{indented}
\item[]\date{\today}
\end{indented}

\begin{abstract}
Helically close-packed states of filaments are common in natural and engineered material systems, ranging from nanoscopic biomolecules 
to macroscopic structural components. While the simplest models of helical close-packing, described by the {\it ideal rope model}, neglect  anisotropy perpendicular to the backbone, physical filaments are often quite far from circular in their cross-section.  Here, we  consider an anisotropic generalization of the ideal rope model and show that cross-section anisotropy has a strongly non-linear impact on the helical close-packing configurations of helical filaments.
We show that the topology and composition of the close-packing landscape depends on the cross-sectional aspect ratio and is characterized by several distinct states of self-contact.  We characterize the local density of these distinct states based on the notion of confinement within a `virtual' cylindrical capillary, and show that states of optimal density vary strongly with the degree of anisotropy.  While isotropic filaments are densest in a straight configuration, any measure of anisotropy leads to helicity of the maximal density state.  We show the maximally dense states exhibit a sequence of transitions in helical geometry and cross-sectional tilt with increasing anisotropy, from {\it spiral tape} to {\it spiral screw} packings.  Furthermore, we show that maximal capillary density saturates in a lower bound for volume fraction of $\pi/4$ in the large-anisotropy, spiral-screw limit.  While cross-sectional anisotropy is well-known to impact the mechanical properties of filaments, our study shows its strong effects to shape the configuration space and packing efficiency of this elementary material motif.

\end{abstract}

%
\vspace{2pc}

\section{\label{sec:intro}Introduction}
Filamentous or fibrous structures are an essential building block of material systems, ranging from the molecular to human scale. At the microscopic end, natural biomolecules~\cite{Neville_1993, Marth_2008, Fratzl_2003} and synthetic polymers~\cite{ZHANG_2022, Paderes_2017, SCShit_2014} form the building blocks of life and of many engineered consumer products, respectively. At macroscopic scale, fibers, wires and ropes form structural materials~\cite{Hearle_1969, Pan_2002, Pan_2014} with diverse applications ranging from  textiles~\cite{BriggsGoode_2011,Schneider_1987,Patti_2023} to tension bearing components in marine mooring lines~\cite{Natarajan_1995, CHAPLIN_1999} and in architecture (e.g. cables in suspension bridges)~\cite{Costello_2012, Krishna_2001}. 

A filamentous material's function, utility, or morphology is inherently intertwined with its \emph{packing}, or space filling arrangement. Entanglements~\cite{Rodney_2005}, knots~\cite{Jawed_2015}, knits~\cite{Singal_2024} and inter-filament orientation~\cite{Gusev_2000} have been shown to influence the mechanical properties of filament assemblies. Precisely folded proteins inherit both function and stability from their ordered arrangement~\cite{Alberts_2003}. Braiding string or tying our shoes requires specific spatio-temporal control of the strand. In each of the above, the packing of the material is essential.


\begin{figure*}
    \centering
    \includegraphics[width=\textwidth]{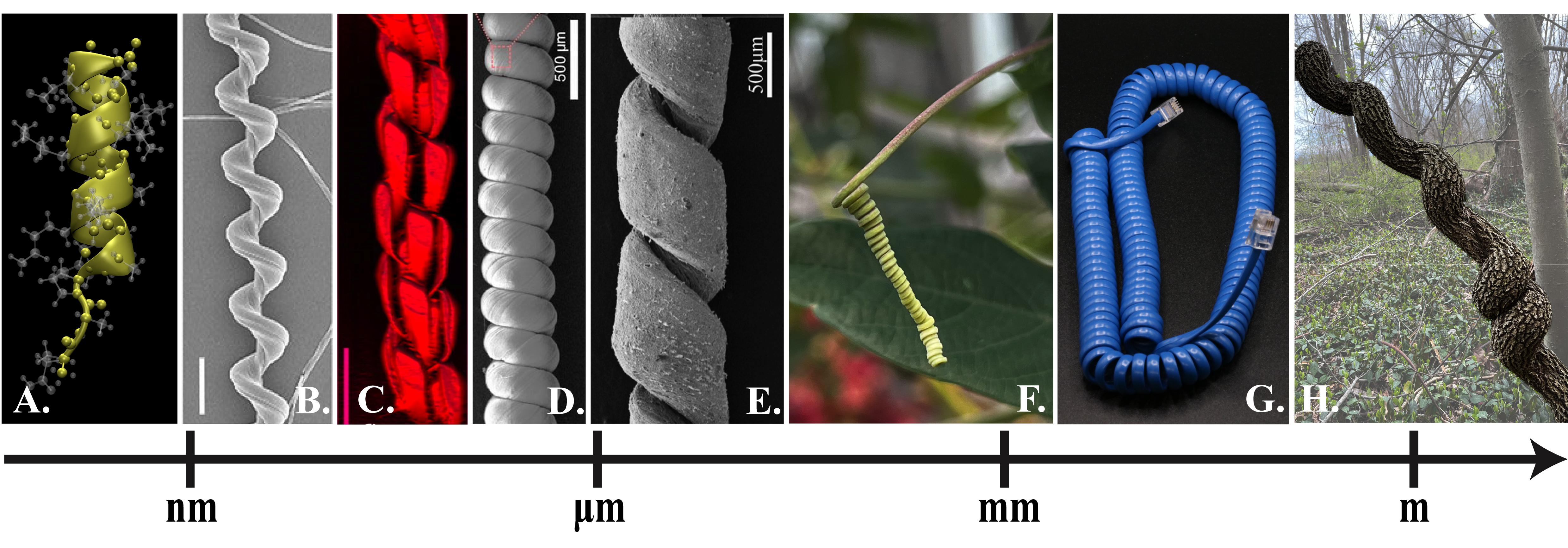}
    \caption{Dense helical coils across length scales. (A.) $\alpha$-helix protein shown using the ``ribbon'' representation; backbone and sidechain atoms rendered in yellow and gray, respectivley (PDB structure from ~\cite{Penin_2000} rendered using VMD~\cite{Humphrey_1996}). 
    (B.) Self-assembled ferrocene-dipeptide ``screw'' nanostructure (scale bar = $1 \mu m$). Adapted with permission from ~\cite{He_2015}. Copyright 2015 American Chemical Society.
    (C.) Nested mesoscale polymer ribbon with tunable photocrease-mediated configuration (scale bar = $50 \mu m$). Adapted from~\cite{Barber_2023}.
    (D.) Twisted helical nanofiber yarn (scale bar = $500 \mu m$). Adapted from~\cite{Guo_2019}.
    (E.) Twisted Graphene Oxide/Alginate hydrogel fiber (scale bar = $500 \mu m$). Adapted with permission from ~\cite{Wang_2020}. Copyright 2020 American Chemical Society.
    (F.) 
    'Blue Bouquet' Passion Flower {\it Passiflora caerulea} tendril self-coiling, photographed at University of Massachusetts Durfee Conservatory.
    (G.) Coiled telephone cable.
    (H.) 
    Oriental Bittersweet ({\it Celastrus orbiculatus}) vine, photographed in Northampton, Massachusetts.}
    \label{fig:fil_examples}
\end{figure*}

Indeed, the packing of material components has been a long-standing and broad interest in mathematics, science, and technology, dating at least as early as Kepler's 17th century conjecture on sphere packing~\cite{Kepler_1611, Aste_2000}, though likely beginning many centuries earlier, especially in the case of mosaics and tilings~\cite{Pickover_2009}. Historically studies on packing have largely focused on arrangements of \emph{discrete}, rigid components (e.g. tiles, spheres, polyhedra). Recently, more attention has been given to the packing of {\it continua}, i.e. soft, deformable and extended objects~\cite{Audoly_2010}, such as flexible sheets~\cite{Cambou_2011, Boue_2006, Vliegenthart_2006}, membranes~\cite{Terasaki_2013, Reich_2019, Tallinen_2016} and filaments~\cite{Grason_2015, Vetter_2015, Weiner_2020, Guerra_2023, Grason_2009}. 
However, despite the natural and technological relevance of filamentous materials, their packing has received considerably less attention, in part due to the complex and vast configuration space accessible to even a single deformable high aspect ratio structure. Due to their flexibility, the packing may be constrained by contact at any location in space and at either local or non-local positions along the filament length \cite{Atkinson_2019}.

Key progress in understanding filament packing geometry is based on what is known as the {\it Ideal Rope} or {\it Ideal Tube} model of packing, which considers space filling configurations based on a filament centerline and a rigid, circular cross-section of uniform diameter $d$ swept-out normal to the backbone curve.  Early applications of this model have been applied to understand the geometry of closed curves in optimally ``tight'' knots and links \cite{Stasiak_1996, Moffatt_1996, Stasiak_text_1998, Cantarella_1998, Gonzalez_1999, Pieranski_1998,VargasLara_2017, Johanns_2021, Klotz_2021, Klotz_2022} 
This model has also been applied to model the complex close contact geometry in multi-filament clasps~\cite{Starostin_2003, Grandgeorge_2021}, as well as plies and bundles ~\cite{Bruss_2012, Starostin_2006, Olsen_AncientRopes_2011, Neukirch_2002,Cajamarca_2014,Crassous_2022},
which have formed a basis for comparison to experimental systems \cite{Ghatak_2005, Panaitescu_2017, Panaitescu_2018, Chopin_2024, Miller_2014, Jawed_2014}. 

A particular focus of the ideal tube model has been on helical close-packing of single filaments, initially motivated as a generic, physical model of condensed helical motifs in biomacromolecules, e.g. $\alpha$-helices, nucleosomes and chromatids \cite{Stasiak_2000, Maritan_2000, Maritan_2008, Olsen_2011, Olsen_2012, Banavar_2023, Schubert_2023, Kamien_2006,Banavar_2003,Chouaieb_2006}. In a seminal work, Przybyl and Pieranski determined the configuration space of helical packing in ideal tubes, solving for the conditions for self-contact that delineate allowed from self-overlapping states in terms of helical radius and pitch of the tubular centerline~\cite{Pieranski_2001}.  Notably, they showed that contact is defined by turn-to-turn stacking of consecutive pitches at large radius, while for small radius contact constraints are local, defined by a minimal curvature radius of the finite-thickness tube. These two contact conditions intersect in configuration space at a particular point, which they consider to be a state of ideal close-packing for helical tubes.  This packing geometry is used to rationalize the stability of similar helical states in models of entropic or osmotically collapsed helical molecules~\cite{Kamien_2006, Hansen-Goos_2007, Poletto_2008}.

While the ideal tube model is a natural starting point for consideration of filamentary close-packing, it is important to recognize that most physical realizations of helical packing present some measure of {\it cross-sectional anisotropy}, deviating by some measure from locally circular shapes.  Fig.~\ref{fig:fil_examples} shows several examples of anisotropic tightly-packed helical structures, across a range of size scales and materials systems.  Some form of cross-sectional asymmetry is present in flexible, densely-coiled filamentous structures at the molecular and meso-engineered systems \cite{Maeda_2016, Barber_2023}, and in a variety of organism structures, including (but certainly not limited to) bacteria \cite{schwan_2013, Rumpel_2008}, curly human hair \cite{Gaines_2023}, plant tendrils and seed pods \cite{Smyth_2016, VanSluys_2021, Silk_2009, Elbaum_2011}, climbing and swimming snakes \cite{Berns_2015, Shine_2011}, elephant trunks \cite{Hu_2022}, and squid and octopus tentacles\cite{Stella_2007, Kier_2012}. Naturally, such filament assemblies lead to the basic question, \emph{how does cross-section geometry influence the states of dense packing in helical filaments}? 

\begin{figure}[h]
    \centering
    \includegraphics[width=5.6in]{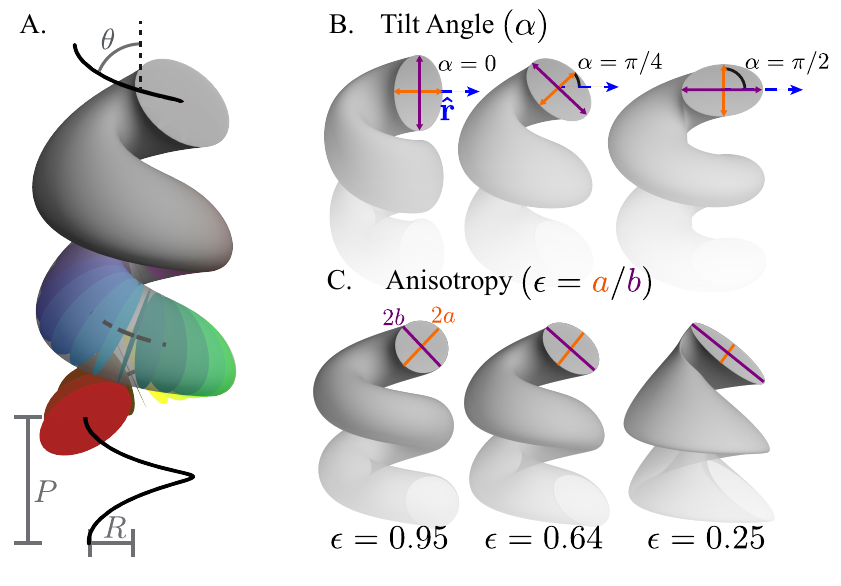}
    \caption{Anisotrpic helical tubes with helical radius $R = 0.6 d_0$ and helical pitch $P = 1.5 d_0$. 
    (A.) Tube surface with helical pitch $P$ and helical radius $R$; together, these parameters define the helical angle, $\theta = \tan^{-1}{\big(}\frac{2 \pi R}{P}{\big )}$.  Everywhere along the helical backbone, the cross-sectional elements are ellipses.
    (B.) Tilt angle ($\alpha$) of the cross-section, where $\alpha$ measures the rotation of the material short axis away from $\hat{\rv}$, see \ref{sec:App_contact}. 
    (C.) cross-section anisotropy. The (shorter) semi-minor elliptical axis has length $a$ while the (longer) semi-major axis has length $b$. $\epsilon$ defines the aspect ratio, $\epsilon = a/b$.
    }
    \label{fig:helical_geometry}
\end{figure}

In this article, we study an extension of the (circular) ideal tube model to include filaments with \emph{elliptical} cross-section, and consider the states of helical packing.  In addition to the radius $R$ and pitch $P$ of the helical backbone, anisotropic tubes are described by two additional parameters:  the elliptical aspect ratio $\epsilon$; and relative orientation, or tilt angle $\alpha$, of the cross-section relative to local bending direction, as illustrated in Fig.~\ref{fig:helical_geometry}.
Like the circular ideal tube model, we model the packing configurations by assuming the filament to be perfectly flexible (i.e. it does not require any energy to bend the filament) but inextensible and unshearable (that is, the cross-section of the tube is everywhere elliptical of specified dimensions).  Based on these assumptions, we fully determine the states of self-contact, which for a given anisotropy $\epsilon$ correspond to a two-dimensional manifold in helical configuration space defined by $R$, $P$ and $\alpha$.  We show that for any measure of anisotropy $(\epsilon \neq 1)$ close contact configurations include a new class of local curvature singularities, corresponding to ``outward folding'' cusps on the surface of low-radius packings.  Moreover, we find that the nature of the self-contact singularity that separates allowed from overlapping configurations is highly sensitive to the degree of anisotropy as well as the orientation of the cross-sectional axis.

To understand conditions for ``optimal'' helical close-packing, we compute a measure of local-density that we call {\it capillary density}, $\phi$, which is simply the occupied volume fraction within a cylindrical volume enclosing the helical tube (of unlimited length)~\cite{Olsen2010TheGG}.  Previous studies of ideal tubes have employed a variety of metrics to define ``maximally-dense'' helical packing, including minimizing the local radius of gyration ~\cite{Pieranski_2001, Maritan_2000} and maximizing excluded volume in a bath of spherical particles~\cite{Kamien_2006}. Our use of capillary density is motivated by the fact that helical packing under cylindrical-like confinement is also prominent in a variety of natural and engineered systems, ranging from DNA coiled within viral phage capsids~\cite{Steven_1997, Phillips_2004}, biopolymers during synthesis~\cite{Hayer-Hartl_2002, Thirumalai_2005}, translocation ~\cite{Maiti_2011, Krantz_2012}, and incorporated within membranes ~\cite{Balusu_2016}, as well as in variety of confined synthetic systems including colloids~\cite{Santang_Dinsmore_2013, zheng_2020_ellipsoids, Hernandez_2021, Wiley_2005,Hutzler_2012,Meseguer_2008,Petukhov_2013}, block copolymers ~\cite{Muller_2020, Russell_2005, Chen_2017}, and liquid crystals ~\cite{Yoon_2014,Yoon_2017,dePablo_2023,Quan_2021}.  Moreover, as we show, the states of optimal capillary density for isotropic versus anisotropic helical tubes are both qualitatively and quantitatively distinct, and capillary close-packing exhibits a strongly non-linear sensitivity to the degree of cross-sectional anisotropy.

While the densest configuration of an isotropic tube under capillary confinement is trivially a straight rod, we show that once cross-sectional isotropy is broken, the densest configurations within \emph{any size} capillary are helical, with a geometry that depends on the level of anisotropy. When the tube is moderately anisotropic, the densest configuration occurs under high confinement (i.e. in a narrow capillary) while increased anisotropy prefers a larger capillary. The densest structures within these two regimes exhibit strikingly different packing motifs both in terms of the helical angle and cross-section orientation; light anisotropy prefers a ``tape-like'' motif reminiscent of Fig.~\ref{fig:fil_examples}E while high anistropy prefers a ``screw-like'' motif reminiscent of Fig.~\ref{fig:fil_examples}B. We determine that these packing regimes crossover when the cross-sectional aspect ratio is approximately $4:1$. While the densest structure (for a given cross-section) never exhibits scrolled or nested packing (i.e. intermediate tilt, like Fig.~\ref{fig:fil_examples}C), we find such structures do optimize density at intermediate levels of confinement, and over a range of confinement that increases with asymmetry.  This analysis reveals that the densest structures for any degree of anisotropy  are {\it bounded}, exhibiting a packing fraction bounded between $1\geq \phi \geq \pi / 4$.

The remainder of this article is as follows. In Section \ref{sec:methods}, we describe the geometric model used to represent our helical tube, the method for finding the boundary between expanded and self-contacting helical states and the \emph{capillary packing fraction}.  In Section \ref{sec:landscapes}, we describe the self-contacting manifold and the packing motifs that constrain different regimes of the configuration space.
In Section \ref{sec:packing}, we assess the capillary packing density of contacting structures, and the dependence on maximally dense structures on cross-section anisotropy.  We discuss the potential implications of these key results in Section \ref{sec:conclusions} for physical and mechanical properties of filamentary systems.

\section{\label{sec:methods}Parameterization and Modes of Contact}
\subsection{3D Helical Geometry}
We model helical configurations of anisotropic tubes using a centerline (or backbone) described by a \emph{circular helix} (i.e. one with constant radius, $R$ and pitch, $P$):

\begin{equation}
    \xv(s) = R \hat{\bf{r}} (s) \pm \frac{P}{\ell} s \hat{z}
\end{equation}
where $\hat{\bf{r}} (s)=\cos (2 \pi s/\ell) \hat{x} + \sin(2 \pi s/\ell)\hat{y} $, $s$ is an arclength coordinate along the backbone and $\ell = \sqrt{4 \pi^2 R^2 + P^2}$ is the arclength per helical turn~\footnote{Note that in this parameterization, the chirality is controlled by the sign of $P$, where $P>0$ corresponds to right-handed coils while $P<0$ refers to left-handed coils. The contact geometry and packing density are, of course, invariant under changes in chirality. In the current manuscript, all schematics shown are left-handed; therefore, listed values of $P$ throughout the text can be take to be negative.}.  Here, we consider a filament backbone of unlimited length, or in effect, neglect contributions to packing from the ends of the structure, appropriate when the total length is much larger than the tube diameter.  The geometry of the helix is independent of backbone position up to screw motions, so the overall configuration can be described by a single ratio, the helical angle, $\theta = \tan^{-1}(\frac{2 \pi R}{P})$ bounded between $\theta = 0$ (a straight line) and $\theta = \pi/2$ (a closed circle).

To describe the material body \emph{anisotropically distributed} around this backbone (Fig.~\ref{fig:helical_geometry}A), we employ an orthonormal material frame, \{$\Tv(s), \hat{\ev}_1(s), \hat{\ev}_2(s)$\} where $\Tv(s)$ is the backbone tangent vector, $\Tv(s) = \xv'(s)= 
\sin \theta \hat{\bf{\phi}} (s) + \cos \theta \hat{z}$ and $\hat{\bf{\phi}} (s)$ is the azimuthal unit vector. We constrain the (undeformable) cross-section to lie within the $\hat{\ev}_1(s)$, $\hat{\ev}_2(s)$ plane, defined in terms of the Frenet-Serret frame (where \{$\Tv(s), \Nv(s), \Bv(s)$\} form an orthonormal triad, see \ref{sec:App_croiss} and Ref~\cite{Kamien_2002})  
\numparts
\begin{eqnarray}
    \hat{\ev}_1(s) = \cos \alpha \Nv(s) + \sin \alpha \Bv(s), \\
    \hat{\ev}_2(s) = \cos \alpha \Bv(s) - \sin \alpha \Nv(s) 
\end{eqnarray}
\endnumparts
Thus, the backbone tangent is normal to the cross-section at all points along the curve. The parameter $\alpha$ describes the ``tilt'' (angle) of the material frame, $\hat{\ev}_1(s)$ and $\hat{\ev}_2(s)$ relative to the normal ($\Nv(s$)) and binormal ($\Bv(s)$) components of the Frenet-Serret frame.  For helices, the normal points toward the pitch axis (i.e. $\Nv(s) = -\hat{\bf{r}}$), and hence, $\alpha$ is easy to visualize as the local orientation of the material frame in the $\hat{\bf{r}}$, $\hat{z}$ plane, as shown in Fig.~\ref{fig:helical_geometry}B.

We model the tubular cross-section as an ellipse with semi-minor axis of length $a$ oriented in the direction of $\hat{\ev}_1(s)$ and semi-major axis of length $b$ oriented in the direction of $\hat{\ev}_2(s)$, see Fig.~\ref{fig:helical_geometry}C. The 3D filament \emph{surface} is then described by
\begin{equation}
\Xv(s, \psi) = \xv(s) + a\cos \psi \hat{\ev}_1(s) + b\sin \psi \hat{\ev}_2(s)
\label{eq: surface}
\end{equation}
where $\psi$ parameterizes the boundary of the elliptical cross-section at a backbone position $s$. 
The material body around the helical backbone is then defined by the aspect ratio of the cross-section, 
\begin{equation}
    \epsilon \equiv a/b
\end{equation}
and the tilt or \emph{rotation of the material frame}, $\alpha$ (Fig.~\ref{fig:helical_geometry}B, C). Intuitively, when $\alpha = 0$, the short material axis is oriented in the direction of curvature (i.e. towards the core of the coil), while $\alpha = \pi/2$ corresponds to the long axis oriented in the curvature direction. Importantly, here we extend the ``Ideal Tube'' model, and therefore treat the filaments as inextensible (fixed length), unshearable (rigid cross-section), and perfectly flexible. In the following, we consider filaments with fixed cross-section area, $A_{\rm fil}$. We define the effective diameter of filaments as, 
\begin{equation}
    d_0 = \sqrt{\frac{4A_{\rm fil}}{\pi}}= 2\sqrt{ab}
\end{equation}
that is independent of $\epsilon$, and we consider all other length scales relative to this microscopic dimension.

\subsection{Sectional geometry}

\begin{figure}[h!]
    \centering
    \includegraphics[width=4.6in]{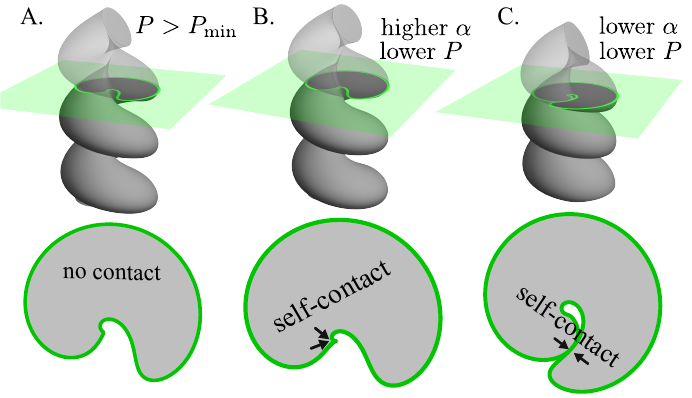}
    \caption{3D and 2D helical configurations for $\epsilon = 0.64$ all with $R/d_0 = 0.5$. (A.) Non-contacting structure with $P > P_{\rm min}$ and $\alpha = 5\pi/16$. (B.) Locally contacting structures obtained from (A.) by \emph{increasing} tilt ($\alpha = 3\pi/8$) and decreasing $P$ until contact, see also Fig.~\ref{fig:SI_cusp} 
    (C.) Non-locally contacting structures obtained from (A.) by \emph{decreasing} tilt ($\alpha = \pi/4$) and decreasing $P$ until contact.}
    \label{fig:modes_of_contact}
\end{figure}

Based on the parameterization of anisotropic tubes defined in eqn.~\ref{eq: surface}, we then determine helical configurations of self-contact by solving for two types of singularities in the tubular surface.  The first correspond to  over-curvature resulting in cusps or creases in the tubular surface, which we denote as {\it local contact}.  The second correspond to overlap of the tube at two distinct surface points, which we denote as {\it non-local contact}

For the purposes of illustrating and analyzing these singular structures it is particular useful to consider the geometry of 2D sections perpendicular to the pitch axis $\hat{z}$, as shown in Fig.~\ref{fig:modes_of_contact}.  We refer to these perpendicular cuts as ``croiss-sections'' due to their crescent-like shape (see Fig.~\ref{fig:modes_of_contact}A). These are defined by solutions to fixed-height equation $\Xv (s, \psi) \cdot \hat{z} = {\rm const.}$, which requires a parameteric relation $s_0(\psi)$ between surface coordinates, from which define the 2D curves
\begin{equation}
\Xv_{\perp}(\psi) = \Xv\big(s_0 (\psi), \psi \big) .
\end{equation} 
For a given helical tube, all croiss-sections are identical up to rigid rotations by screw symmetry.  Fig.~\ref{fig:modes_of_contact}A shows an example for a non-contacting helical configuration.  Fig.~\ref{fig:modes_of_contact}B shows an example of a local contact singularity for higher tilt and minimal pitch, which appears as an ``outward'' cusp in $\Xv_{\perp}(\psi)$ for this example.  Fig.~\ref{fig:modes_of_contact}C shows an example of a non-local contact singularity for lower tilt and minimal pitch, which appears as two distinct locations on the curve $\Xv_{\perp}(\psi)$ meeting at single point in space.

\subsection{Local Contact}
Local contact occurs at a single $\psi$ along the tube surface where the surface fails to be smooth; this discontinuity manifests as either an inward or outward crease in the $\Xv(s, \psi)$ surface (or a cusp of the same "direction" in the $\Xv_{\perp}(\psi)$ slice, see Fig.~\ref{fig:modes_of_contact}B, Fig.~\ref{fig:SI_cusp}). Operationally, these singularities can be selected by locations where the \emph{differential surface element} of the tube vanishes. Using the surface metric, $\rm g$, the magnitude of the surface normal can be expressed as 
\begin{equation}
    \sqrt{\rm det(g)} = | \partial_s \Xv(s, \psi) \times \partial_\psi \Xv(s, \psi) | 
\end{equation}
which yields the surface area of a differential tube element as $dA = \sqrt{\rm det(g)} \, ds \, d\psi$.  

The conditions for local contact are defined by the solutions to $\rm det(g) = 0 $ (see eqns.~\ref{eqn:Metric1} --~\ref{eqn:Metric3} in \ref{sec:App_contact}).  Notably, the Ideal Tube case ($\epsilon =1$) corresponds to a singular limit of these equations in which local-contact takes the form on an inward crease, at a minimum pitch derived by Przybyl and Pieranski \cite{Pieranski_2001} 
\begin{equation}
    \label{eqn:PP_Plocal}
    P_{\rm local} = 2 \pi \sqrt{\frac{d_0}{2} R - {R}^2}, \ \ \ \ {\rm for} \ \epsilon =1
\end{equation}
More generally for $\epsilon \neq 1$ there are two solutions, corresponding to local singularities at distinct surface locations, given by 
\begin{eqnarray}
    \label{eqn:Plocal_anisoA}
    P_{\rm local(-)} =  2 \pi \sqrt{a \cos \alpha R - {R}^2 }, \ \ {\rm for} \ \ \epsilon \neq 1 \\
    \label{eqn:Plocal_anisoB}
    P_{\rm local(+)} = 2 \pi \sqrt{b \sin \alpha R - {R}^2 }, \ \  {\rm for} \ \ \epsilon \neq 1 
\end{eqnarray}
which also varies with orientation of the tubular cross-section .

While these local-contact solutions share similarities with the isotropic (Ideal Tube) case, there are notable differences in \emph{which} elliptical axis sets the curvature constraint. The first solution (eqn.~\ref{eqn:Plocal_anisoA}) is constrained by the cross-section semi-minor axis ($a$) and is maximized when this length is oriented in the radial ($\hat{\rv}$) direction (i.e. $\alpha = 0$). On the other head, the second solution (eqn.~\ref{eqn:Plocal_anisoB}) is constrained by the cross-section semi-major axis ($b$) and is maximized when the cross-section is rotated by $\alpha = \pi / 2$, such that this length is oriented in the $\hat{\rv}$ direction. These two regimes have distinct geometrical structure and consequences of constraining close-packed geometries, discussed in Section \ref{sec:landscapes}.

\subsection{Non-local Contact}
The tube surface makes contact non-locally when two distinct elements of the croiss-section occupy the same point in space (Fig.~\ref{fig:modes_of_contact}C), which like the local-contact case, implies contact along a 1D helical curve where two parts of the same tube meet along its length.

In searching for non-local contact, we compute the \emph{distance of closest approach} between these distinct points along the tube surface. When the cross-section is isotropic, it is sufficient to calculate the distance between points on the backbone and require that they be separated by a distance equal to the tube diameter $d_0$~\cite{Pieranski_2001, Gonzalez_1999}. Because of the asymmetry in the cross-section, we must instead compute distances between {\it surface elements}. Our approach employs several coupled criteria which determine points of distance of closest approach along the tube surface (e.g. that the separation is normal to the tube surface at {\it both} points ) then vary the configuration parameters (i.e. $R$, $P$ and $\alpha$) until this closest distance is zero (i.e. the points are in contact, see \ref{sec:App_contact}). In practice, we fixed combinations of $R$ and $\alpha$, and vary $P$, numerically solving for $P_{\rm non-local}(R, \alpha)$.  

While it is straightforward to see that for sufficiently large $P$ any combination of $R$ and $\alpha$ will be embeddable (i.e. non-overlapping), in general, it is not a priori known what type of contact first limits the reduction of pitch at a given $R$ and $\alpha$ (i.e. local or non-local contact).  We determine the pitch at which the tube first makes contact at a given given $R$ and $\alpha$, i.e. minimum non-overlapping pitch, as the largest pitch at which the tube either makes local or non-local contact,  $P_{\rm min}(R, \alpha) = {\rm Max}\big[P_{\rm local(\pm)}(R, \alpha),P_{\rm non-local}(R, \alpha)\big]$.

\subsection{Capillary Confinement}
\begin{figure}[h!]
    \centering
    \includegraphics[width=\textwidth]{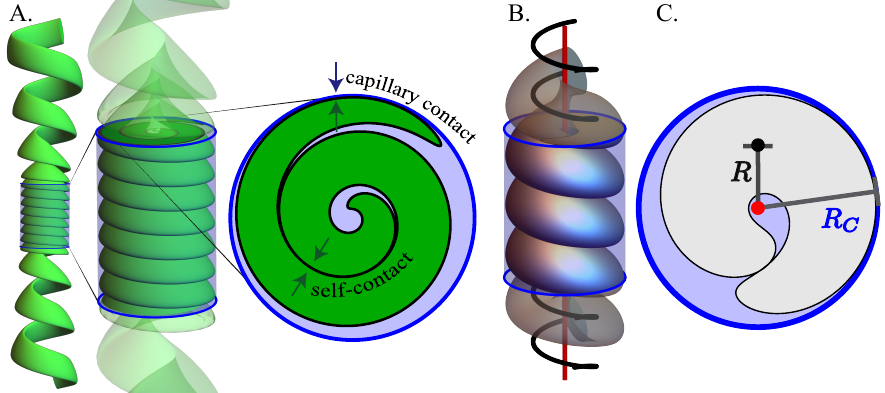}
    \caption{Dense capillary packing. (A.) Hypothetical filament constrained by capillary- and self-contact. Schematic of uniform filament confined in a (B.) 3D and (C.) 2D cylindrical pore of radius $R_C$}
    \label{fig:capillary_schematic}
\end{figure}

For illustrative purposes, consider a expanded, loosely coiled filament. If this filament is compressed until self-contact is made, then a confining capillary is constricted around assembly until the capillary is in contact with the filament surface. In a sense, the capillary contact constrains the structure from dilating radially, while the filament self-contact prevents further radial compression; this conceptual paradigm is presented in Fig~\ref{fig:capillary_schematic}A.

 Operationally, we numerically determine the smallest cylinder (with radius $R_C$) which could confine the structure, shown in Fig~\ref{fig:capillary_schematic}B,C.  Because each planar section of the packing is identical up to a rigid rotation about the pitch axis, the occupied volume fraction in the capillary is identical to the occupied area fraction of the croiss-section within the circular boundary of the cylinder. Because the filament cross-sections are unshearable, and by Pappus's theorem the enclosed volume per unit backbone length is constant~\cite{Goodman_1069}, it follows that $A_{{\rm croiss}}P = A_{{\rm fil}} \ell$, where $A_{{\rm croiss}}$ is the area enclosed by the cross-section normal to the pitch axis and $A_{{\rm fil}}$ is the area of (elliptical) cross-section normal to the backbone. Using the fact that $A_{{\rm croiss}}/A_{{\rm fil}} = \sec \theta$ we compute the capillary packing fraction as

\begin{equation}
    \label{eqn:phi}
    \phi = \frac{A_{{\rm croiss}}}{\pi {R_C}^2} = \frac{d_0^2 \sec \theta}{4 R_C^2}
\end{equation}

where $R_C={\rm max}_\psi |{\bf X}_\perp (\psi)|$. Rationale for the use of the minimal size capillary is discussed in~\ref{sec:App_phi_vs_RC}.

 \section{\label{sec:landscapes}Close-packed configuration landscapes}
In this section, we analyze the manifold self-contacting helical geometries, specifically, analyzing the minimal non-overlapping pitch $P_{\rm min}(R,\alpha)$for for a given helical backbone radius and cross-sectional tilt $\alpha$ for a sequence of cross-sectional anisotropy $\epsilon$.

\subsection{\label{sec:isotropics}Isotropic tubes}

\begin{figure}[h!]
    \centering
    \includegraphics[width=\textwidth]{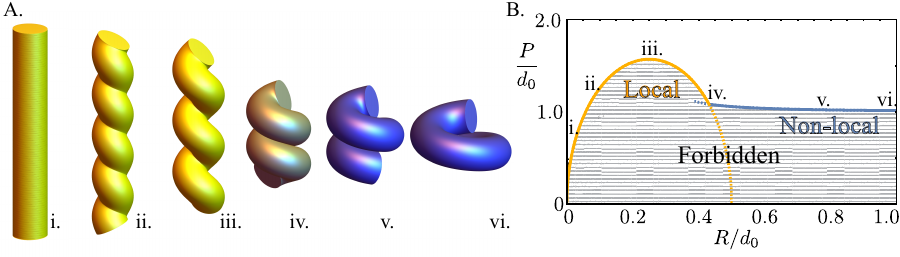}
    \caption{Packing landscape for isotropic tube, $\epsilon = 1$, originally calculated in ref.~\cite{Pieranski_2001}. (A.) Maximal pitch structures for increasing values of $R/d_0$. (B.) Minimal pitch contact hull indicating the limiting mode of contact. }
    \label{fig:iso_p}
\end{figure}

When the cross-section is isotropic (e.g. circular, $a=b$), the packing landscape is determined only by helical radius (e.g. it is invariant to tilt). Przybyl and Pieranski observed that contact is delineated into two regimes~\cite{Pieranski_2001}; when $R$ is small, the structure is curvature-limited and makes self-contact locally, or along an inward crease at the center of helix (analogous to overbending a garden hose) as illustrated in Fig.~\ref{fig:iso_p}A i.-iii.  Packing at large $R$ is limited by (non-local) contact between successive turns of the tube (analogous to coil stacking of a garden hose), illustrated in Fig.~\ref{fig:iso_p}A v.-vi. Here, the single boundary between local and non-local contact occurs at $R/d_0 \simeq 0.431$ and $P/d_0 \simeq 1.083$ (Fig.~\ref{fig:iso_p}A iv.); a single local maximum in $P_{\rm min}/d_0= \pi/2$ occurs at $R/d_0 = 0.25$. Together, these contact criteria generate the minimum pitch bounding hull, reproduced in the style of \cite{Pieranski_2001} in Fig.~\ref{fig:iso_p}B, where structures below $P_{\rm min}(R)$ are forbidden (due to self-overlap) and structures above $P_{\rm min}(R)$ are said to be ``expanded''.  

\subsection{\label{sec:highEps}Slightly anisotropic tubes}

\begin{figure}[h!]
    \centering
    \includegraphics[width=\textwidth]{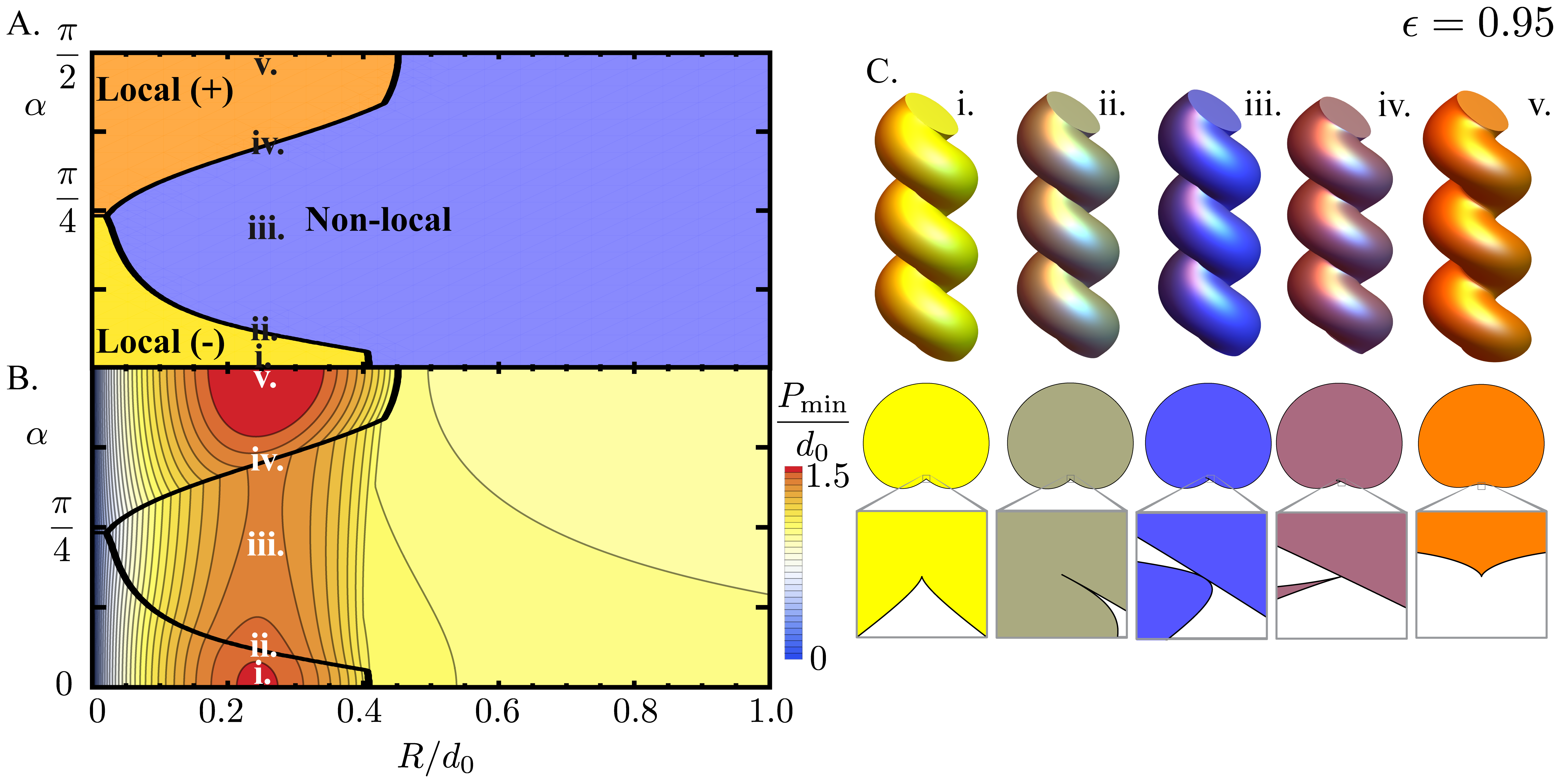}
    \caption{Packing landscape for $\epsilon = 0.95$. (A.) Contact type phase diagram and (B.) minimal pitch contact hull. (C.) Maximal pitch structures for increasing values of $\alpha$.}
    \label{fig:e095_p}
\end{figure}

With apparently any perturbation to the cross-section shape (or at least with as small as $0.5 \%$ deviation, $\epsilon = 0.995$), we find that 
the modes of contact change significantly; notably, now with a vital dependence on cross-section tilt.  These large changes for weak anisotropy illustrate that the isotropic tube case ($\epsilon \to 1$) is in fact a singular limit of the more general variable anisotropy behavior.  As shown in Fig.~\ref{fig:e095_p}B, the large $R$ behavior is largely unchanged (albeit with a slight tilt-bias which intuitively grants large $\alpha$ structures the ability to pack at lower pitch). However, when $R$ is small in the regime corresponding to local contact of the isotropic filament, we observe three distinct contact regimes, shown in Fig.~\ref{fig:e095_p}A. Two of these regimes are determined by local contact corresponding to solutions of vanishing area element of the metric tensor(eqn. \ref{eqn:Plocal_anisoA} and \ref{eqn:Plocal_anisoB}), however these represent distinct types of contact, in particular driven by the long material axis ($b$, at high $\alpha$) and the short material axis ($a$, at low $\alpha$). Surprisingly, the low $R$ - intermediate $\alpha$ behavior exhibits non-local contact; this seemingly stems from variation in tube-surface curvature (resulting from cross-section anisotropy) thereby smoothing what was otherwise an inward local kink. This packing paradigm allows for a slight softening of the previous global maximum in $P_{\rm min}$; in other words, non-local contact at intermediate tilt allows for a lower pitch configuration. While these changes are challenging to recognize visually at the 3D scale (structures along the $P_{\rm min}(R, \alpha)$ ridge are shown in Fig.~\ref{fig:e095_p}C), their modes of contact are distinct when magnified; we note that the low-$\alpha$ local contact creases ``inward'' in the structure, manifesting as an \emph{intruding groove} (and hence labeled ``Local (-)'') while the high-$\alpha$ local contact creases ``outward'', manifesting as a \emph{protruding seam} (and hence labeled ``Local (+)''). It is also important to note that despite the minuscule differences in contact geometry and fairly modest shifts in the values of $P_{\rm min}$ at a given $R$, the boundaries between distinct contact topologies exhibit a large shift in then $R$-$\alpha$ plane, even when the cross-section anisotropy is small. These shifts in the contact manifold become even more pronounced as the tubular cross-section becomes more anisotropic, as we discuss next.

\subsection{\label{sec:lowEps}Highly anisotropic tubes}
When the cross-section is symmetric (with respect to tilt), the packing landscapes are also symmetric (see Fig.~\ref{fig:iso_p}B). Introducing asymmetry in the cross-section results in asymmetry in the landscape; the Local (+) regime occupies slightly more area than the Local (-) regime in Fig.~\ref{fig:e095_p}A and the corresponding maxima in Fig.~\ref{fig:e095_p}B is more pronounced.

\begin{figure}[h!]
    \centering
    \includegraphics[width=\textwidth]{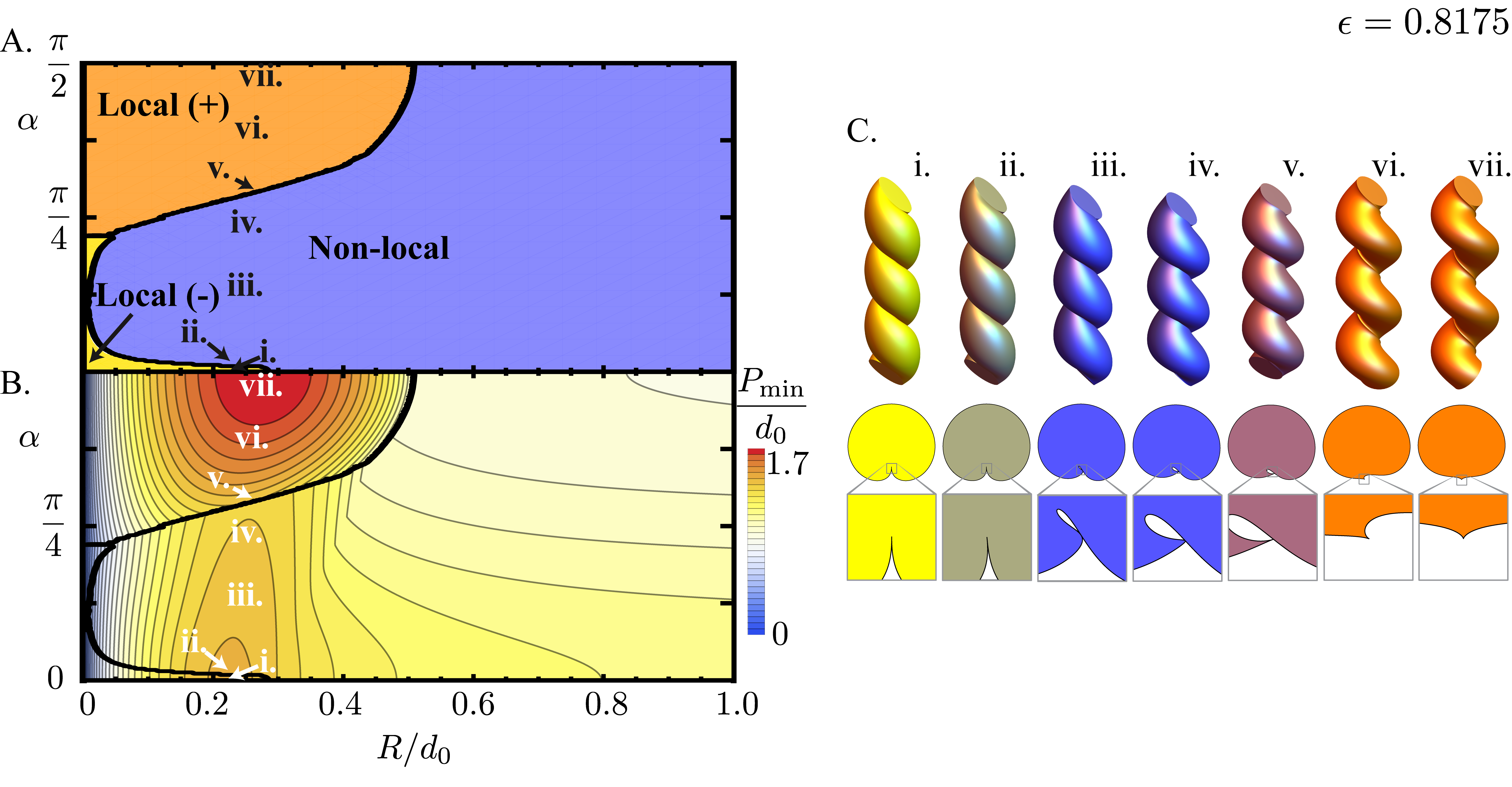}
    \caption{Packing landscape for $\epsilon = 0.8175$. (A.) Contact type phase diagram and (B.) minimal pitch contact hull. (C.) Maximal pitch structures for increasing values of $\alpha$.}
    \label{fig:e08175_p}
\end{figure}

As the filament becomes more anisotropic (see Fig.~\ref{fig:e08175_p}A,B  for $\epsilon = 0.8175$), the asymmetry between + and $-$ local contact grows. With decreasing $\epsilon$, the Local (+) and Non-local regions become more prominent while the Local (-) shrinks, and the differences in $P_{\rm min}$ between large- and small-$\alpha$ become significant. The minimal pitch also exhibits a stronger tilt-dependence; while two maxima separated by a saddle still dominated the low-$R$ behavior, the magnitude of the maxima (corresponding to structures Fig.~\ref{fig:e08175_p}C.i and C.vii now differ by approximately 20 percent (with $P_{\rm min} \simeq 1.42 d_0$ and $1.73 d_0$, respectively). At this scale, the 3D structures exhibit some perceptible asymmetry; structures with intermediate tilt (Fig.~\ref{fig:e08175_p}C.ii-vi) display uneven protrusions in the radial direction, while structures constrained by Local (+) curvature (Fig.~\ref{fig:e08175_p}C.v-vii) display a noticeable seam.

When the cross-section is increasingly asymmetric (below $\epsilon \simeq 4/5$), the Local (-) branch ceases to constrain the packing manifold, at which point, the entire low-tilt regime is limited by non-local turn-to-turn contact exclusively. When the cross-section has \emph{at least} this degree of anisotropy, it is not possible to construct a viable configuration that is more curved than the short material length scale would allow because when the structure is in the low-tilt regime, the curvature of the filament is primarily in the direction of the short material axis.  There simply is not enough material in this direction to curve such that a kink arises, while in the pitch-direction, the cross-section is wide which obstructs larger regions of turn-to-turn configurations.

\begin{figure}[h!]
    \centering
    \includegraphics[width=\textwidth]{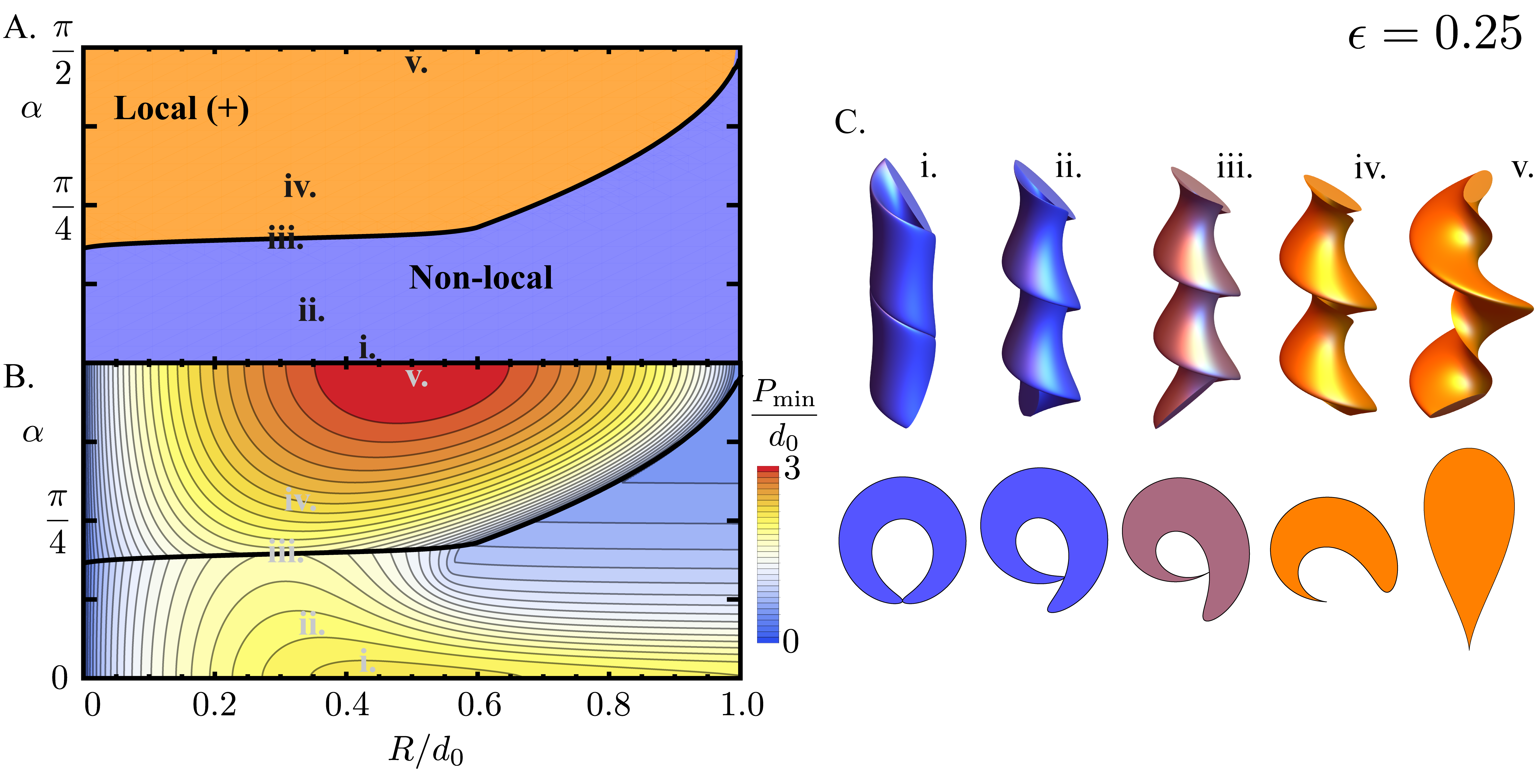}
    \caption{Packing landscape for $\epsilon = 0.25$. (A.) Contact type phase diagram and (B.) minimal pitch contact hull. (C.) Maximal pitch structures for increasing values of $\alpha$.   }
    \label{fig:e025_p}
\end{figure}

An example of the ``two-phase'' contact topology for highly-anisotropic tubes ($\epsilon = 0.25$) is shown in Fig.~\ref{fig:e025_p}. As the filament becomes more anisotriopic, the global maxima of $P_{\rm min}(R, \alpha)$ at $\alpha = \pi/2$ become more pronounced (Fig.~\ref{fig:e025_p}B). The secondary maxima at low-$R$ for $\alpha = 0$ persists, despite the change between modes of contact (from Local (-) to Non-local), though the feature continues to softens relative to the global maximum (Fig.~\ref{fig:e025_p}C.i and C.v have $P_{\rm min} \simeq 2.25 d_0$ and $\simeq 3.15 d_0$ at $\alpha = 0$ and $\alpha = \pi/2$, respectively). In this asymmetry regime, these extrema remain separated by a saddle which spans intermediate tilt; however, the span is not smooth (i.e. $\partial P_{\rm min} / \partial \alpha$ is discontinuous), owing to the boundary between modes of contact. In these significantly asymmetric cases, the corresponding structures exhibit exotic morphologies; a selection that spans the $P_{\rm min}$ saddle are shown in Fig.~\ref{fig:e025_p}C. In particular, the global maxima of $P_{\rm min}$ Fig.~\ref{fig:e025_p}C.v., corresponding to Local (+) contact at small-$R$, $\alpha = \pi/2$ presents a relatively large $P_{\rm min}$ characterized by elaborate outward cusps at the high curvature edge of the cross-section. This behavior is in stark contrast to the Non-local large-$R$, $\alpha = \pi/2$ packing, in which $P_{\rm min}$ is minimal (i.e. $P_{\rm min}(R\to \infty,\alpha=\pi/2) \rightarrow 2a$, or ``small-axis stacking''). Intuitively, as the radius of curvature is compressed beyond the width of cross-section in radial direction, the packing must expand correspondingly into the $\hat{z}$ dimension (resulting in larger $P_{\rm min}$). 

  
Przybyl and Pieranki considered the transition point between local and non-local contact branches to correspond to a state of optimally dense helical packing. In the next section, we consider the density of these much richer close-packed landscapes, and in particular, the behavior in proximity to packing motif coexistence branches. Additionally, a more complete analysis of contacts along the coexistence branch are provided in \ref{sec:App_phi_vs_RC}. 

\section{\label{sec:packing}Capillary packing}

\begin{figure*}
    \centering
    \includegraphics[width=\textwidth]{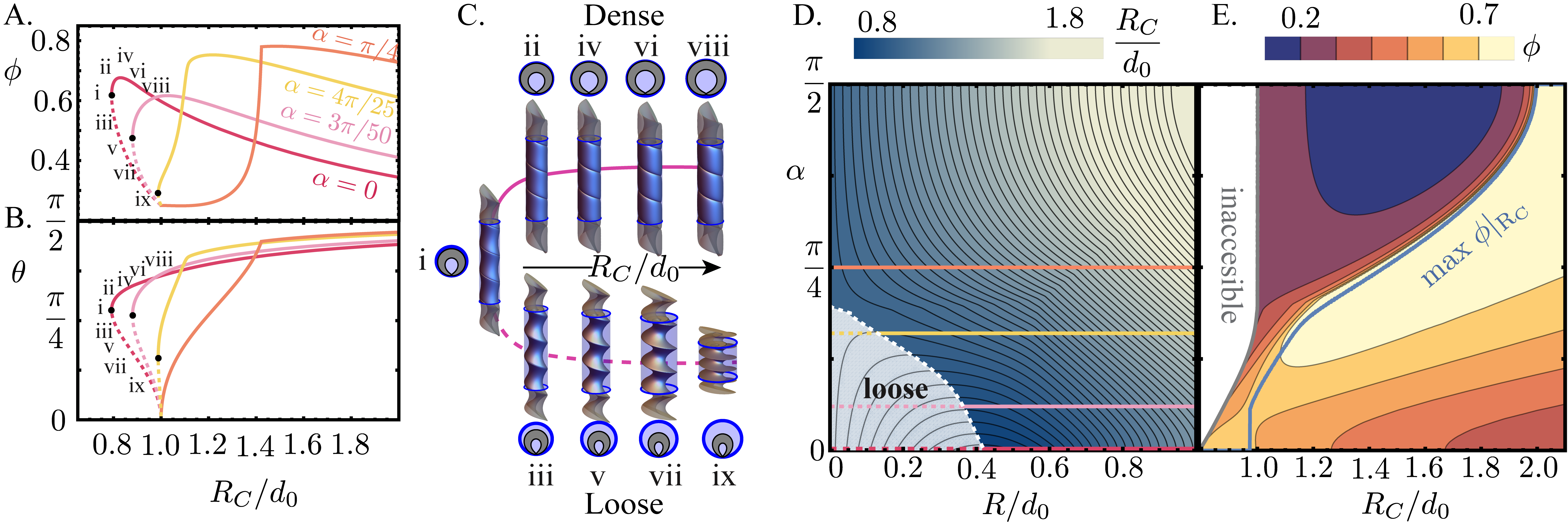}
    \caption{Multiplicity of states within a given capillary of $R_C$ shown for $\epsilon = 0.25.$. Traces of (A.) density and (B.) helical angle for \emph{fixed} tilt angle as a function of confining radius. 
    (C.) Dense and loose structures within the same capillary $R_C/d_0= 0.79$ (i.), $0.84$ (ii, iii), $0.89$ (iv, v), $0.94$ (vi, vii), $0.99$ (viii, ix).
    (D.) $R_C$ evaluated along the contact hull; the region at low-$R$ low-$\alpha$ corresponds to bifurcated regions,
    (E.) Landscape of maximum with variable $R_C$ and $\alpha$. A plot of the maximal density at fixed $R_C$ is overlaid.
    }
    \label{fig:metastability}
\end{figure*}
Having found that helical geometry of filament self-contact is strongly sensitive to the anisotropy of the cross-section, we now consider its effects on capillary packing fraction, $\phi$.

In this section, we first briefly summarize the approach to computing optimally dense configurations. We then describe features of the packing density manifold and the underlying structural motifs as a function of the size of the confining capillary, $R_C$. Finally, we analyze and compare the overall densest configurations as a function of tube aspect ratio, $\epsilon$ and classify the distinct packing motifs selected for variable filament asymmetry and confinement.

\subsection{Measuring local density by capillary confinement}
In the prior sections, it was most convenient to parameterize packing in terms of the purely geometric parameters that describe the configuration, in particular, the radius $R$ and pitch $P$ of the tubular ``centerline''. This description is sufficient for capillary confinement of isotropic tubes ($\epsilon =1$) because the tightest capillary radius for a given helical radius is simply $R_C = R+d_0/2$.  For anisotropic tubes ($\epsilon \neq 1$), the mapping between helical centerline radius $R$ and capillary radius, is a more complex and non-linear function $R_C(R, P,\alpha)$.  In practice, this is computed from the maximal radial distance of the tubular surface from the pitch axis, i.e. $R_C={\rm max}_\psi |{\bf X}_\perp (\psi)|$ for a given $R, P$ and $\alpha$.  That is, the capillary radius is the smallest possible to enclose the helical filament, or other words, the minimal possible value of $R_C$ is chosen so as to maximize the packing fraction for that $R, P$ and $\alpha$.

For all close-packed (i.e. self-contacting) structures for a given $\epsilon$, that is, all combinations of $R$, $\alpha$, and the resulting minimal pitch $P_{\rm min}$, we calculate $\phi(R, \alpha)$ and $R_C(R, \alpha)$\footnote{Note that by constraining our optimization to close-packed structures, $\phi(R, \alpha)$ and $R_C(R, \alpha)$ have an implicit dependence on $P_{\rm min}$, where $P_{\rm min}$ = $P_{\rm min}(R, \alpha)$ }. By inverting the mapping $R_C(R, \alpha)$ (to yield $R(R_C, \alpha)$), we analyze the relationship between confinement size for a close-packed structure and its packing fraction.  
As an example, we show several traces of $\phi(R_C, \alpha)$ for fixed $\alpha$ and $\epsilon = 0.25$ in Fig.~\ref{fig:metastability}A-B. Importantly, when both $\alpha$ and $R_C$ are small, $\phi(R_C)$ is {\it multivalued} since there are multiple $R$ values with the same $\alpha$ that \emph{perfectly} fit within a specific capillary (see schematic examples in Fig.~\ref{fig:metastability}C.).  

As we are presently interested in the densest packing, we take only the higher $\phi_C$ configuration in this multi-valued region in our analysis, neglecting the features of the ``loose'' (larger pitch) branch of capillary confinement.

For $\epsilon = 0.25$ the map between helical radius and capillary radius is shown in Fig. ~\ref{fig:metastability}D, where the ``loose'' structures occupy a region at low-$R$ low-$\alpha$.  The boundary of this region (white dashed line in Fig. ~\ref{fig:metastability}) corresponds to the minimal capillary radius for a given $\alpha$ in this regime. 
For all confined packings along this boundary between single- and multi-valued helical radius, we observe helical geometry ($\theta > 0$, marked with black dots in Fig.~\ref{fig:metastability}B), which is not the case for larger tilt and lower asymmetry, where the tightest capillary radius is achieved for $R \to 0$.  For example, at values of tilt larger than the bifurcation region for $\epsilon = 0.25$, $\alpha \gtrsim 0.6$, the minimal capillary size is independent of $\alpha$; all structures exhibit a minimal $R_C$ governed by $\theta = 0$ ``rod-like'' packing, in which $R_C$ is exactly the semi-major cross-sectional axis, $b$. Regardless of aspect ratio (and tilt), these rod-like packings exhibit a density $\phi = A_{{\rm fil}}/\pi {R_C}^2 = \pi a b / \pi b^2 = \epsilon$.

Taken together, these effects lead to complex landscapes of capillary packing density as a function of cross-section tilt and capillary radius, as shown for example for $\epsilon =0.25$ in Fig.~\ref{fig:metastability}E.  Here, $\phi$ spans $0.8 \gtrsim \phi \gtrsim 0.1$. Within the parameter range shown, $\phi$ exhibits a minimum when $R_C/d_0 \simeq 1.5$, $\alpha = \pi/2$; structures in this neighborhood correspond to high-$P_{\rm min}$ Local (+) structures, like Fig.~\ref{fig:e025_p}C.v. Of course, these configurations only correspond to a local minima; $\phi \rightarrow 0$ as $R_C/d_0 \rightarrow \infty$, resulting from the center of the coil opening.

In the following, we next consider $\alpha$ as a configurational degree of freedom (like $R$ and $P$) and determine the optimal tilt and helical geometry for dense packing of a tube with a given anisotropy $\epsilon$ and in a given capillary radius $R_C$.  For the example of $\epsilon =0.25$ shown in Fig.~\ref{fig:metastability}E, the optimal $\alpha$ value for each given $R_C$ is shown as blue curve, which transitions from $\alpha = 0$ for small $R_C$ to $\alpha = \pi/2$ for large $R_C$, a generic feature for all $\epsilon$ as we discuss below.

\subsection{Variable-Radius Capillary Packing}

\begin{figure*}
    \centering
    \includegraphics[width=\linewidth]{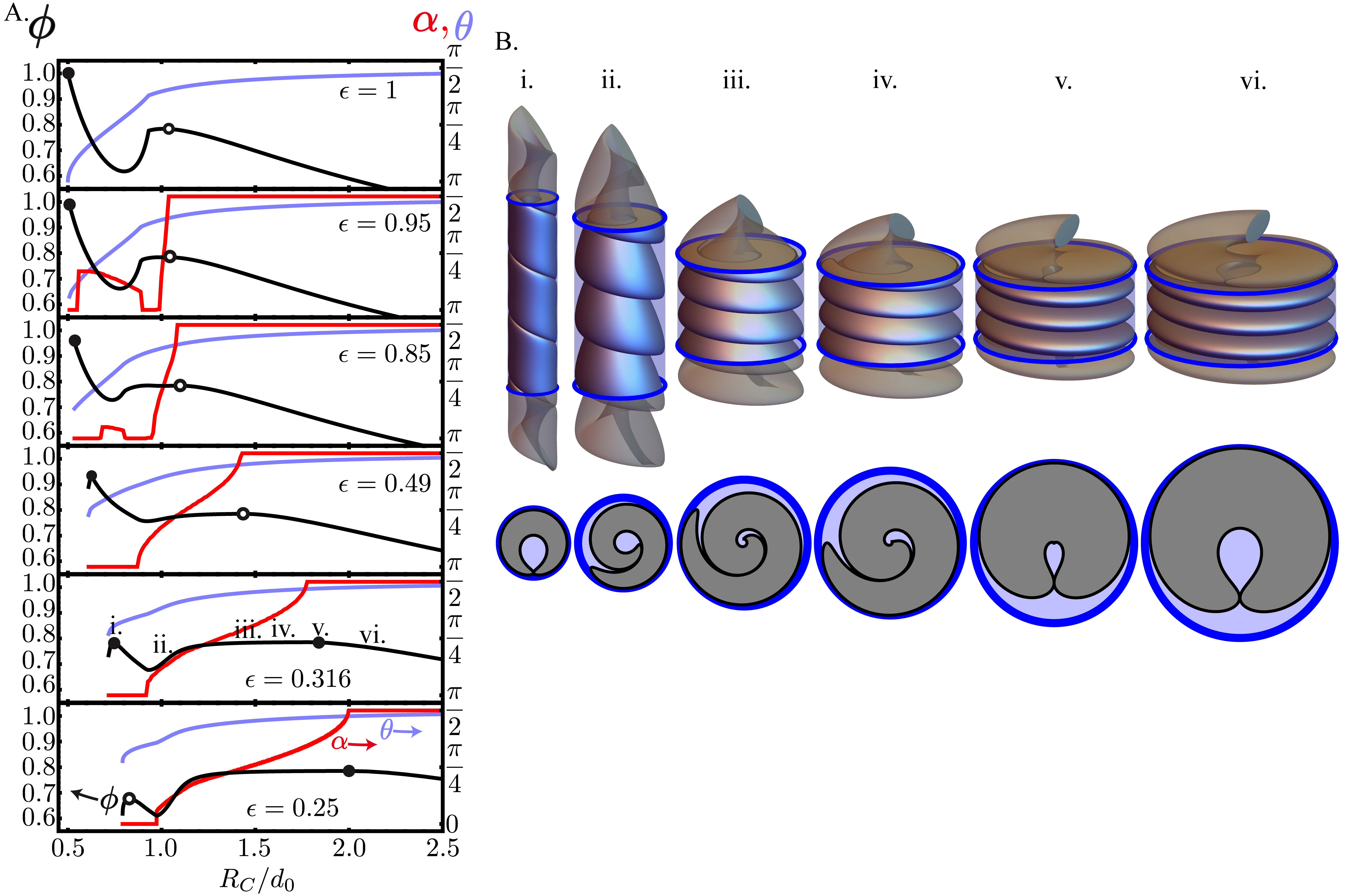}
    \caption{Densest configurations within variable sized capillaries. 
    (A.) Maximal density ($\phi$, black, left axis), tilt angle ($\alpha$, red, right axis), helical angle ($\theta$, blue, right axis) within capillary $R_C/d_0$ for selected values of $\epsilon$. Global maxima of $\phi$ marked with solid dots; local maxima marked with open dots.
    (B.) Structures for $\epsilon = 0.316$ within increasing $R_C/d_0$, displaying increasing $\alpha$ and $\theta$. 
    }
    \label{fig:phi_vs_R}
\end{figure*}

Fig.~\ref{fig:phi_vs_R}A shows plots of the optimal capillary packing density and configurations as function of capillary radius ($R_C$) and for a sequence of increasing cross-sectional asymmetry. For all aspect ratios of the filament, packing fraction $\phi$ exhibits a non-monotonic dependence on the confining radius. When the tube is isotropic ($\epsilon = 1$), the packing exhibits four extrema (Fig.~\ref{fig:phi_vs_R}A,top):
\begin{enumerate}
    \item when the capillary is the same size as the tube, the tube perfectly packs within the capillary ($\phi = 1$). We consider this to be a ``rod-like'' packing motif.
    
    \item when the capillary size is intermediate to the tube radius and tube diameter, the structure is helical with lower packing fraction, corresponding to high-$P_{\rm min}$ local contact structures.
    
    \item when the capillary size is comparable to the tube diameter, the packing fraction reaches a local maxima of $\phi = \pi/4$; based on the density, we consider this to be a ``toroid-like'' packing motif \footnote{Toroid-like packing exhibits a packing density $\phi = \pi/4$, equivalent to that of a horn-torus within a cylinder of radius and height equal to the tube diameter; interestingly, the helical capillary exhibits the same ratio of size (a cylinder of radius equal to the tube diameter)}.

    \item when the $R_C$ is larger than the tube diameter, the core of the helical structure begins to open, leading to a monotonic decrease in density with increasing capillary size ($\phi \rightarrow 0$ with $R_C \rightarrow \infty$).   
    
\end{enumerate}

The helical angle $(\theta = \tan^{-1}(\frac{2 \pi R}{P}))$ exhibits a monotonic dependence on $R_C$. When the filament is isotropic and the cylinder is in the small limit ($R_C = d_0/2$), the tube is straight (i.e, $R = 0$) thus $\theta = 0$. In the large limit ($R_C \rightarrow \infty$), the tube approaches a toriodal geometry (i.e, a helix with $P \rightarrow 0$) thus $\theta = \pi/2$.


Helical packing of asymmetric tubes ($\epsilon \neq 1$) leads to new features of the capillary dependent packing, shown in Fig.~\ref{fig:phi_vs_R}A.  With any degree of anisotropy, we observe that the minimal capillary radius no longer corresponds to a local maxima in density; this maxima shifts to larger capillaries and decreases in density with increasing anisotropy, first very slightly for low anistropy (e.g. $\epsilon = 0.95$) and then more obviously with larger anisotropy $(\epsilon = 0.49)$ The minimum helical angle (which occurs within the smallest possible capillary radius) increases with anisotropy, never limiting to the isotropic value of $\theta = 0$; in other words, the straight, non-helical tube is \emph{never} a density maximizer (for an aniostropic tube), regardless of size of the capillary. For any degree of cross-sectional anisotropy, capillary density favor helical over straight configurations.

Relative to small-$R_C$, the large-capillary behavior is qualitatively insensitive to anisotropy; the density exhibits a local maxima when the capillary radius is equal to twice the semi-major tube axis. Naturally, this value of $R_C$ grows as with increasingly tubes anisotropy.

Given a tube anisotropy, optimal capillary density requires strongly non-linear variation in the tilt angle ($\alpha$) of the cross-section. Near maxima of $\phi(R_C)$, we observe extremes in the tilt angle. When $R_C$ is small, the cross-section is oriented with the short axis pointing (inward) in the curvature direction ($\alpha = 0$). This general motif, observed at the low $R_C$ density maxima, is reminiscent of a {\it coiled tape}, on for example, bicycle handlebars, and shown for $\epsilon = 0.316$ in Fig.~\ref{fig:phi_vs_R}B.i.  When $R_C$ is larger, the dense structures exhibit a cross-section with the long axis oriented (inward) in the curvature direction ($\alpha = \pi/2$; this motif is reminiscent of the threads on a screw, see Fig.~\ref{fig:phi_vs_R}B.v; the onset of this regime universally exhibits the toroidal density, $\phi = \pi/4$.

For large anisotropies ($\epsilon \lesssim 4/5$, these states are connected at intermediate $R_C$ via a monotonic increase in the tilt angle, visualized in Fig.~\ref{fig:phi_vs_R}B.ii-iv. The width of the transition broadens with anisotropy, revealing an increasing number of dense, intermediate-tilt states that exhibit a nested or ``scroll''-like packing. Interestingly, these states with large but sub-maximal tilt exhibit a packing density nearly degenerate to the $\phi = \pi/4$ toroidal packing.  When the tube is moderately anistropic, we also note the existence of a set of low-$R_C$ nested states; these structures correspond to a slight preference for non-local contact over Local (-) contact, and persists until the Local (-) regime no longer constrains the close-packing landscape ($\epsilon \simeq 0.81$). Given the slight amount of anisotropy, these configurations are structurally similar to the untilted structures; the packing benefit and resulting structural changes are slight in this case.

We mark the locations of low-$R_C$ (tape-like) and high-$R_C$ (screw-like) density maxima with black markers in Fig.~\ref{fig:phi_vs_R}A. For each value of $\epsilon$, the global maximum is marked with a solid dot, while the secondary maximum is marked with an open dot. When $\epsilon$ is high (i.e. nearly isotropic), the low-$R_C$ maxima is most dense, and when $\epsilon$ is small (i.e. large anisotropy), the high-$R_C$ maxima is most dense. 
This transition in the geometry of maximally dense packing is examined in more detail in the next section. 


\subsection{Optimal Capillary Packing}

\begin{figure}
    \centering
    \includegraphics[width=\linewidth]{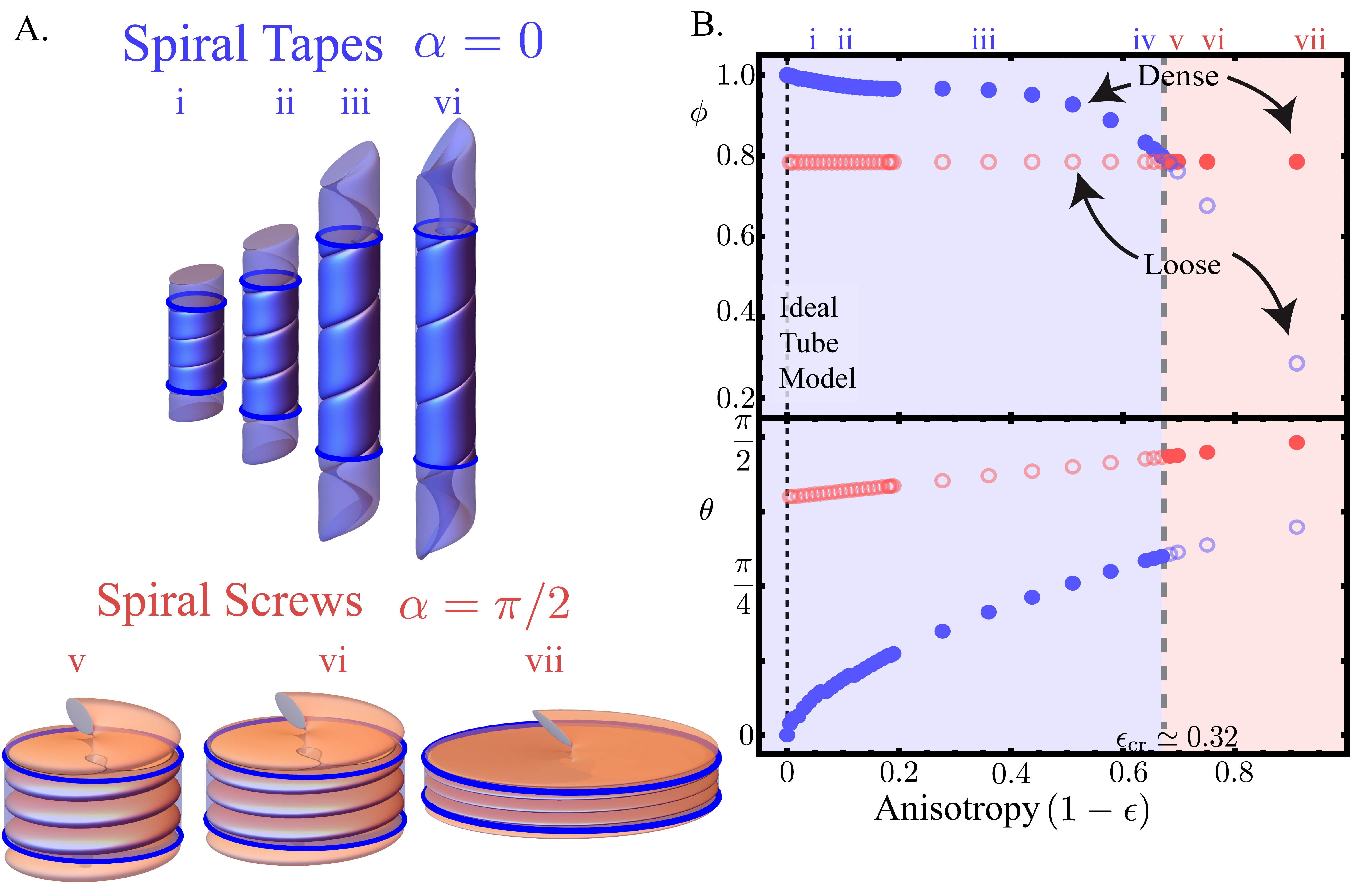}
    \caption{Competing dense packing motifs. 
    (A.) Densest configurations for selected values of $\epsilon$.
    (B. top) Density ($\phi$) global (``\emph{dense}'') and local (``\emph{loose}'') maxima. Global maxima are shown as solid dots while submaximal packings are shown as open dots. 
    (B. bottom) Helical angle ($\theta$) for densest configurations}
    \label{fig:phi_vs_epsilon}
\end{figure}

As shown in Fig.~\ref{fig:phi_vs_R}, $\phi(R_C)$ is characterized by two local maxima, one at low-$R_C$ with small $\theta$ and a second at large-$R_C$ and large $\theta$, which are respectively closer to nearly-straight versus ``gardenhose-like'' coiling.  In Fig.~\ref{fig:phi_vs_epsilon}, we track each of these two localy dense states (i.e the $\phi$ maxima in Fig.~\ref{fig:phi_vs_R}). Again, global maxima (``dense'') are marked with a solid point while the secondary maxima (``loose'') are marked with an open point.
When the tube is either isotropic or only moderately elliptical, the density of the small-$R_C$ ($\alpha = 0$, low $\theta$) state is higher than that of the large-$R_C$ ($\alpha = \pi/2$, high $\theta$) state. However, the small-$R_C$ density decreases with anisotropy, while the large-$R_C$ density remains constant; at a critical aspect ratio, $\epsilon_{\rm cr} \simeq 0.316$, the densest state transitions between solution branches. This dependence is intuitive to understand by simply considering the capillary fraction of a straight tube, $\phi(\theta =0)=\epsilon$, which obviously decreases due to the inefficiently of filling a circular capillary with an elliptic cross-section.  Hence, secondary ``gardenhose'' maxima eventually overtakes the nearly-straight state in density.  This transition between packing motifs shows a characteristic tilt- and helical angle dependence; the densest high-$\epsilon$ structures pack with $\alpha=0$ and low $\theta$ (or a  \emph{steep} incline) while the low-$\epsilon$ structures with $\alpha=\pi/2$ orientation and high $\theta$ (or a \emph{gradual} incline); the globally optimal configurations for assorted values of $\epsilon$ are shown in Fig.~\ref{fig:phi_vs_epsilon}A. These competing motifs are shown at approximately the transition point, $\epsilon_{\rm cr} \simeq 0.316$, in Fig.~\ref{fig:phi_vs_R}B.i and B.v. Curiously, this transition occurs when the aspect ratio is $\simeq \pi^{-1}$, at which point the density of both the tape-like and screw-like structure exhibit $\phi = \pi/4$.



\begin{figure}[h!]
    \centering
    \includegraphics[width=\textwidth]{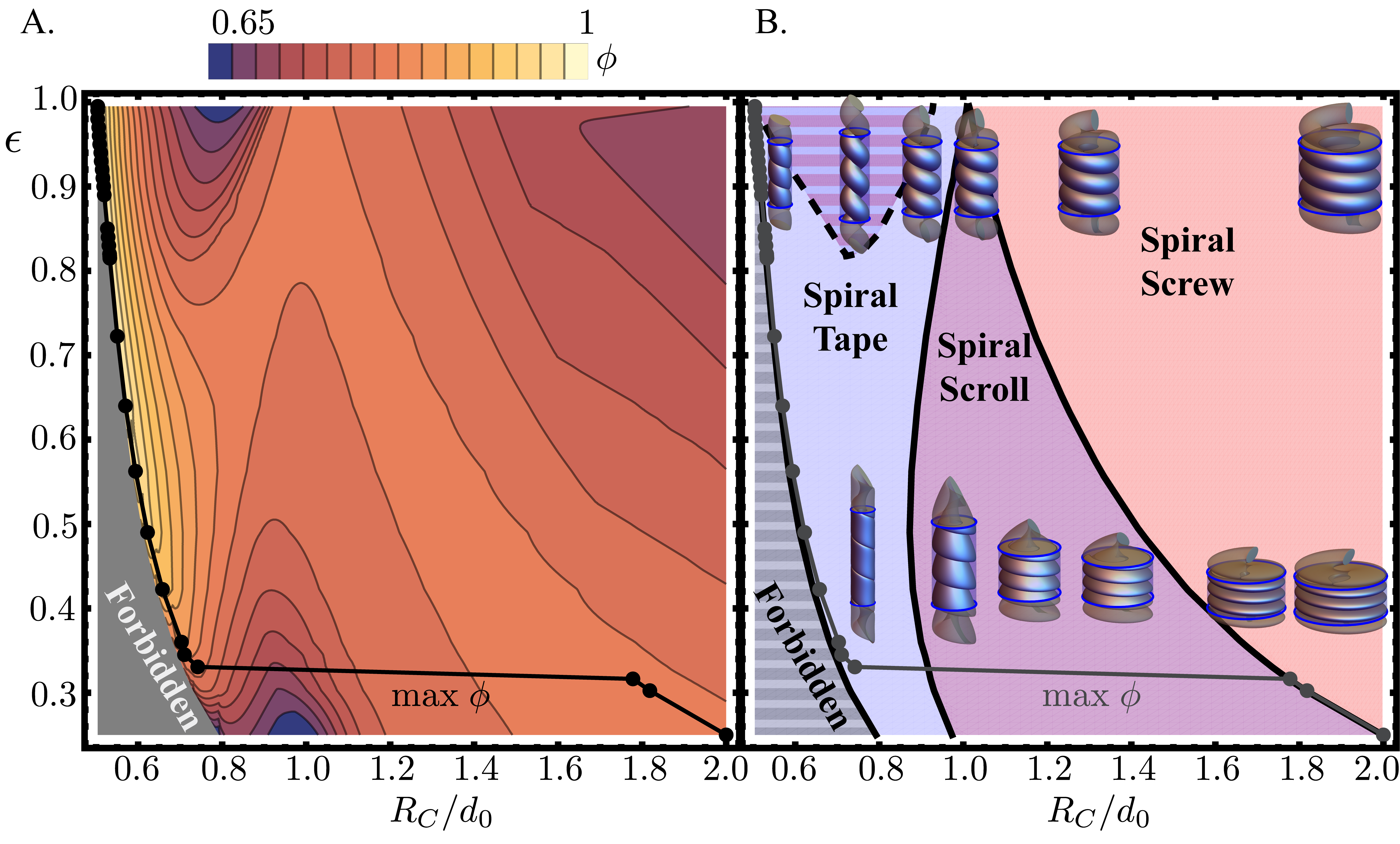}
    \caption{Anisotropy-confinement landscapes. (A.) Density of packed filaments with aspect ratio $\epsilon$ constrained to capillary with radius $R_C/d_0$. (B.) Configurational phase diagram consists of spiral tapes ($\alpha = 0$), spiral scrolls (($\pi/2 > \alpha > 0$)), and spiral screws ($\alpha = \pi/2$).   }
    \label{fig:epsilon_vs_Rc}
\end{figure}


\section{\label{sec:conclusions}Discussion and concluding remarks}



In this article, we have inventoried the dense-packing parameter space of helical tubes with cross-sections of various aspect ratio (or \emph{flatness}).  We find that once cross-section symmetry is broken, the tube configuration space is constrained by novel and non-trivial variants of the local (curvature-limited) and non-local (turn-to-turn) contact for isotropic filaments reported in \cite{Pieranski_2001}. This complex spectrum of closed-packing geometries, and their non-linear dependence on anisotropy has important implications for the ``local density'' of helical close-packing, measured in terms of capillary packing density.
While the densest structure is straight for isotropic tubes, once cross-section symmetry is broken, the densest configuration is either tape-like ($\alpha = 0$) when the tube is mildly anisotropic or screw-like ($\alpha = \pi/2$) if the tube is sufficiently anisotropic.  Importantly, the density of the weakly-anisotropic regime decreases with $\epsilon$ while the latter is independent of $\epsilon$ (and displays a packing fraction of $\phi = \pi/4$, equivalent to that of a horn torus confined to a tight cylinder). Therefore, the densest configuration \emph{for any filament} is bounded $1\geq \phi \geq \pi / 4$, assuming an optimal capillary. Regardless of the morphology, the ``locally dense'' packing motif required an increasingly larger capillary as the tube became more anisotropic. These trends of optimal density are summarized in Fig.~\ref{fig:epsilon_vs_Rc}A. Based on our highest aspect ratio studied here ($\epsilon = 0.09$), it seems reasonable that the low $\epsilon$ preference for $\alpha = \pi/2$ packing should hold in the limit of thin ribbon- or sheet-like helices (i.e. $\epsilon \rightarrow 0$); perhaps, not surprisingly, these packings seem to be abundant in biological membranes, including in the endoplasmic reticulum~\cite{Terasaki_2013} and plant (photosynthetic) thylakoid membranes~\cite{Reich_2019}. 

The strong influence of cross-sectional anisotropy on \emph{optimal} close-packing, raises additional question about how (density and morphology) evolves under sub-optimal confinement. Though the optimal structures always required extremal (tape-like with $\alpha = 0$ or screw-like with $\alpha = \pi/2$) tilt, intermediate levels of confinement often preferred intermediately tilted, nested scroll-like configurations, as shown in the morphological phase diagram in Fig.~\ref{fig:epsilon_vs_Rc}B (see \ref{sec:App_contact} for a discussion of the small tape/scroll region at low $R_C$ and high $\epsilon$). Many of these nested configurations (especially at large $\alpha$) exhibit a density nearly degenerate to the screw-like structure (see the soft gradients in Fig.~\ref{fig:epsilon_vs_Rc}A at low $\epsilon$, intermediate $R_C$). Further, the nested states occupy a larger portion of confinement-parameter space as the cross-section becomes more asymmetric. Given a driving force for densification (like some applied compressive force), these findings hint at the potential to control the assembly morphology via the degree of confinement.

In considering the anisotropic filament packing geometry, we account for more complex material \emph{shape}, however we have thus far restricted ourselves to very simple mechanical descriptions, namely, neglecting considerations of bend and twist elasticity of filaments. A more complete mechanical model should take into elastic costs of helical deformations, and specifically account for  elastic anisotropies that result from the geometric anisotropies~\cite{Audoly_2010, Palmer_2020}.
Based on simple Kirchoff rod theory, one should expect (for a filament composed of an isotropic elastic medium) bending is stiffer in the {\it wide} axis relative to the {\it narrow} axis by a factor of $\epsilon^{-2}.$  On these grounds, and assuming that filaments are straight and untwised in their rest configuration, it is reasonable to expect that the $\alpha = \pi/2$ configurations would impose a higher elastic cost (assuming a straight rest configuration) than $\alpha =0$ packings.  This suggests a basic antagonism between geometries that favor optimal dense packing and those that require low elastic energy.  
Further, we can imagine a class of situations for which the rest state of filaments is helical~\cite{Lazarus_2013a, Miller_2014, Chouaieb_2004, DOMOKOS_2005, Wang_2020}, 
but the geometry of close-packing may indeed frustrate the minimal elastic energy.  One example of this scenario is likely to be elastically programmed microfilaments that target inaccessible values of helical pitch and radius~\cite{Barber_2023}.  Notably, accounting for both density and elasticity in this situation is likely to direct the structures towards the tilted, nested structures shown in Fig.~\ref{fig:fil_examples}C.

Of course, the current question of dense packings underlies a range of other interesting phenomena, including the mechanical response of helical structures under an applied load, the impact of incompatible coiling (e.g. geometric frustration arising from an inaccessible elastic ground state) on the resulting structure, multi-component (and size-disperse) interacting systems including bundles \cite{Grason_2015, Hall_2016} and arrays \cite{Moed_2024}.  In particular, it remains to be explored how the mapping between packing of filaments in helically-twisted structures and packing on non-Euclidean surfaces~\cite{Bruss_2012, Grason_2015} sheds useful light on optimal states of packing anisotropic filaments.  For example, for the present model of helical filament configurations with elliptical cross-sectional asymmetry,  it can be expected that local and global geometric constraints can be understood by packing of geodesic ellipses on positive curvature surfaces~\cite{Andraz_2022}.

\ack

The authors are grateful to K. T. Sullivan for helpful comments on this manuscript. Numerical calculations were carried out on the Unity Cluster, a collaborative, multi-institutional high-performance computing cluster managed by the UMass Amherst Research Computing and Data team. This research was supported by a National Science Foundation Graduate Research Fellowship awarded to B.R.G.

\section*{Conflict of Interest}
The authors declare no conflict of interest.

\section*{Data Availability}
The analysis code that was used for this study is openly available in UMass Amherst ScholarWorks at \url{https://doi.org/10.7275/gr0k-fv13}.

\appendix

\section{\label{sec:App_croiss}Surface and Sectional Geometry}

\subsection{Frenet-Serret Formulas}

To describe the material body \emph{anisotropically inflated} around the helical backbone (Fig.~\ref{fig:helical_geometry}A), we employ the orthonormal (geometric) Frenet-Serret frame, composed of the tangent, normal, and binormal triad, \{$\Tv, \Nv, \Bv$\}:

\begin{equation}
    \Tv(s) = \xv'(s) = \sin \theta \hat{\bf{\phi}}(s) + \cos \theta \hat{z}(s)
\end{equation}

\begin{equation}
    \Nv(s) = \Tv'(s) = - \hat{\bf{r}}(s) 
\end{equation}

\begin{equation}
    \Bv(s) = \Tv(s) \times \Nv(s) = - \cos \theta \hat{\bf{\phi}}(s) + \sin \theta \hat{z}(s)
\end{equation}

where \{$\hat{\bf{r}}, \hat{\bf{\phi}}, \hat{z}\}$ are the canonical cylindrical coordinate basis vectors and $\tan\theta = 2 \pi R / P$:
\begin{equation}
    \hat{\bf{r}} = \cos (2 \pi s / \ell) \hat{x} +  \sin (2 \pi s / \ell) \hat{y}
\end{equation}

\begin{equation}
    \hat{\bf{\phi}} = - \sin (2 \pi s / \ell) \hat{x} +  \cos (2 \pi s / \ell) \hat{y}
\end{equation}

\subsection{2D ``Croiss-section'' Derivation}
We select the contents of the tube that lie in the $\hat{x}$, $\hat{y}$ plane at fixed height, $z_0$
\begin{equation}
\Xv(s, \psi) \cdot \hat{z} = z_0  
\end{equation}

\begin{equation}
\frac{P}{2 \pi l} s + a \cos \psi \sin \alpha \sin \theta + b \sin \psi \cos \alpha \sin \theta = z_0  
\end{equation}

This constraint allows us to eliminate one of the two parametric variable by yielding the relationship between backbone ($s$) and surface ($\psi$) coordinate (that satisfies the planar constraint).

\begin{equation}
s_0(\psi) = -\frac{2 \pi \ell \sin \theta}{P}(a \cos \psi \sin \alpha + b \sin \psi \cos \alpha + z_0) 
\end{equation}


By substitution of this constraint, the full 3D model reduces to a 2D one-coordinate function (see Fig.~\ref{fig:modes_of_contact}), where the (in-plane) boundary of the tube is given by

\begin{equation}
\Xv_{\perp}(\psi) = \Xv\big(s_0 (\psi), \psi \big)
\end{equation}


And the 2D tangent of the surface is given by

\begin{equation}
    \Tv_{\perp}(\psi) = \frac{\Xv_{\perp}'(\psi)}{|\Xv_{\perp}'(\psi)|}
\end{equation}

\section{\label{sec:App_contact}Determination of Self-Contact}

\subsection{Non-local Contact}

To determine the distance between any two in-plane points, we define the separation vector, 

\begin{equation}
    {\bf{\Delta_{\perp}}} (\psi_1, \psi_2) = \Xv_{\perp}(\psi_1) - \Xv_{\perp}(\psi_2)
\end{equation}

The distance of closest approach is defined between surface coordinates $(\psi_1$ and $\psi_2)$ when ${\bf{\Delta_{\perp}}} (\psi_1, \psi_2)$ is orthogonal to both surface tangents, $\Tv_{\perp}(\psi_1)$ and $\Tv_{\perp}(\psi_2)$ (or, equivalently, $\Tv_{\perp}(\psi_1) \parallel \Tv_{\perp}(\psi_2)$. In determining instances of non-local contact, we seek to simultaneously satisfy $\Tv_{\perp}(\psi_1) \parallel \Tv_{\perp}(\psi_2)$, ${\bf{\Delta_{\perp}}} (\psi_1, \psi_2) \rightarrow 0$, and $\psi_1 \neq \psi_2$. 

Recalling that the $\Xv_{\perp}(\psi)$ croiss-section plane has $\hat{z}$ as a normal, we define three inequalities to capture the distance of closest approach 
\begin{equation}
{\big |} \Tv_{\perp}(\psi_1) \cdot {\bf {\Delta_{\perp}}} (\psi_1, \psi_2) {\big |}^2 \geq 0
\end{equation}

\begin{equation}
{\Big |} {\big [}\Tv_{\perp}(\psi_1) \times  \Tv_{\perp}(\psi_2) {\big ]} \cdot \hat{z} {\Big |}^2 \geq 0
\end{equation}

\begin{equation}
{\Big |} {\big [} \Tv_{\perp}(\psi_2) \times  {\Delta_{\perp}} (\psi_1, \psi_2){\big ]} \cdot \hat{z}{\Big |}^2 \geq 0
\end{equation}


When the first two quantities saturate $0$, $\psi_1$ and  $\psi_2$ correspond to a location of distance of closest approach; these points are in the neighborhood of non-local contact (given a sufficiently small $P$). The third quantity extracts the separation distance between the points; when this quantity \emph{also} equals $0$, non-local contact occurs. We numerically solve this coupled system for saturation of all three constraints (in terms of the contact pitch, $P_{\rm non-local}$ and the surface contact coordinates, $\psi_1$ and  $\psi_2$) for every combination of $R$, $\alpha$, and $\epsilon$. In this procedure, we impose a constraint that $\psi_1 \neq \psi_2$; the angular constraints baked into our inequalities then maintain that $\psi_1$ and  $\psi_2$ are in fact non-local along the tube surface. 

\subsection{Local Contact}

Local contact occurs at a single location along the tube surface where the surface fails to be smooth; this discontinuity manifests as a crease in the surface or cusp singularity in the in-plane representation. Operationally, these singularities can be selected by locations where the \emph{ differential surface area} of the tube vanishes, corresponding to the surface being multivalued.

Using the surface metric, $\rm g$, the magnitude of the surface normal can be expressed as 
\begin{equation}
    \sqrt{\rm det(g)} = | \partial_s \Xv(s, \psi) \times \partial_\psi \Xv(s, \psi) | 
\end{equation}

which yields the surface area of a differential element as
\begin{equation}
    dA = \sqrt{\rm det(g)} \, ds \, d\psi 
\end{equation}

We select vanishing area elements via solutions to $\rm det(g) = 0$, which yields a sum of three squares (corresponding to component for each of the orthonormal basis components), ${c_1}^2(\psi) + {c_2}^2(\psi) + {c_3}^2(\psi) = 0$. 


\begin{equation}
    \label{eqn:Metric1}
        \fl {c_1}^2(\psi) =  b^2 \cos \psi \sin \psi - a^2 \cos \psi \sin \psi \\
\end{equation}

\begin{equation}
    \label{eqn:Metric2}
        \fl {c_2}^2(\psi) =  ab \cos^2\psi \cos\alpha \sin\theta - b^2 \cos \psi \sin \psi \sin \alpha \sin \theta  -\ell b \cos \psi \\ 
\end{equation}

\begin{equation}
    \label{eqn:Metric3}
       \fl {c_3}^2(\psi) = - ab \sin^2\psi \sin\alpha \sin\theta + 
        a^2 \cos \psi \sin \psi \cos \alpha \sin \theta - \ell a \sin \psi
\end{equation}

Of course, satisfying ${c_1}^2(\psi) + {c_2}^2(\psi) + {c_3}^2(\psi) = 0$ requires that each term equal zero. 

Importantly, the first condition, ${c_1}^2$, exhibits a singular dependence on the aspect ratio, $\epsilon$; when the cross-section is circular ($\epsilon = 1$), the condition is automatically satisfied for any value of $\psi$ and the minimal pitch ($P_{\rm local}$ takes the form of a helix \emph{``limited by it's curvature''} presented by Przybyl and Pieranski \cite{Pieranski_2001}, see eqn.~\ref{eqn:PP_Plocal}.

When $\epsilon < 1$, satisfying the initial constraint (${c_1}^2(\psi)=0$)  requires that the product $\cos\psi\sin\psi = 0$; naturally, this requires that either $\cos\psi$ or $\sin\psi = 0$ (i.e. $\psi = {0, \pi/2, \pi, 3\pi/2, 2\pi}$). Substitution of these constraints into ${c_2}^2(\psi)$ and ${c_3}^2(\psi)$ yields two physical solutions for $P_{\rm local}$, see eqn.~\ref{eqn:Plocal_anisoA} and \ref{eqn:Plocal_anisoB}:


These solutions present as ellipses in the $(R, P)$ plane, making it trivial to identify their center and vertices. The ellipses are described by


\begin{equation}
    \frac{{P}^2}{ (a \pi  \cos\alpha)^2} + \frac{(R - \frac{1}{2}a\cos\alpha)^2}{(\frac{1}{2} a\cos\alpha)^2} = 1
\end{equation}

and

\begin{equation}
    \frac{{P}^2}{ (b \pi  \sin\alpha)^2} + \frac{(R - \frac{1}{2}b \sin\alpha)^2}{(\frac{1}{2}b \sin\alpha)^2} = 1
\end{equation}

\begin{figure}[h!]
    \centering
    \includegraphics[width=\textwidth]{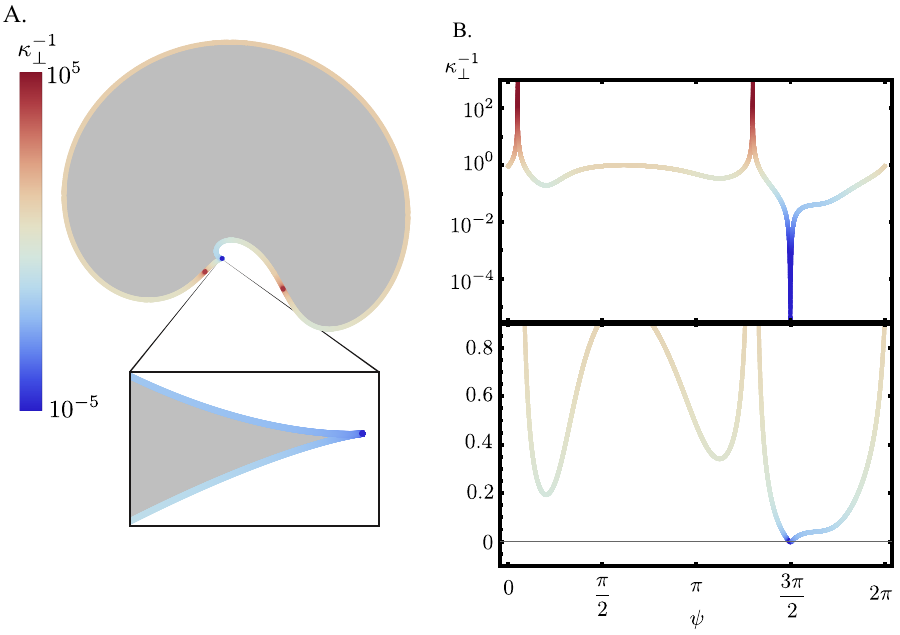}
    \caption{Radius of curvature ($\kappa_{\perp}^{-1}$) along $\Xv_{\perp}(\psi)$ corresponding to Fig.~\ref{fig:modes_of_contact}B(top); In this structure, $\kappa_{\perp}^{-1}$ spans 0 (cusp) to $\infty$ (inflection); for visualization the color bar scales logarithmically and was truncated to saturate at $\kappa_{\perp}^{-1} = 10^{-5}$ (dark blue) and $\kappa_{\perp}^{-1} = 10^{5}$ (dark red).}
    \label{fig:SI_cusp}
\end{figure}

Perhaps more intuitively (or at least more graphically), this contact feature can also be detected numerically by a diverging in-plane curvature; the geometric significance of both operations are the same. In Fig~\ref{fig:SI_cusp}, the in-plane radius of curvature, $\kappa_{\perp}^{-1} = | \Xv_{\perp}'(\psi)|^{3/2}/ |\Xv_{\perp}'(\psi) \times \Xv_{\perp}''(\psi)|$ is shown for the locally self-contacting structure presented in Fig.~\ref{fig:modes_of_contact}B.

\subsection{The Local - Non-local Boundary}

\begin{figure*}
    \centering
    \includegraphics[width=0.8\textwidth]{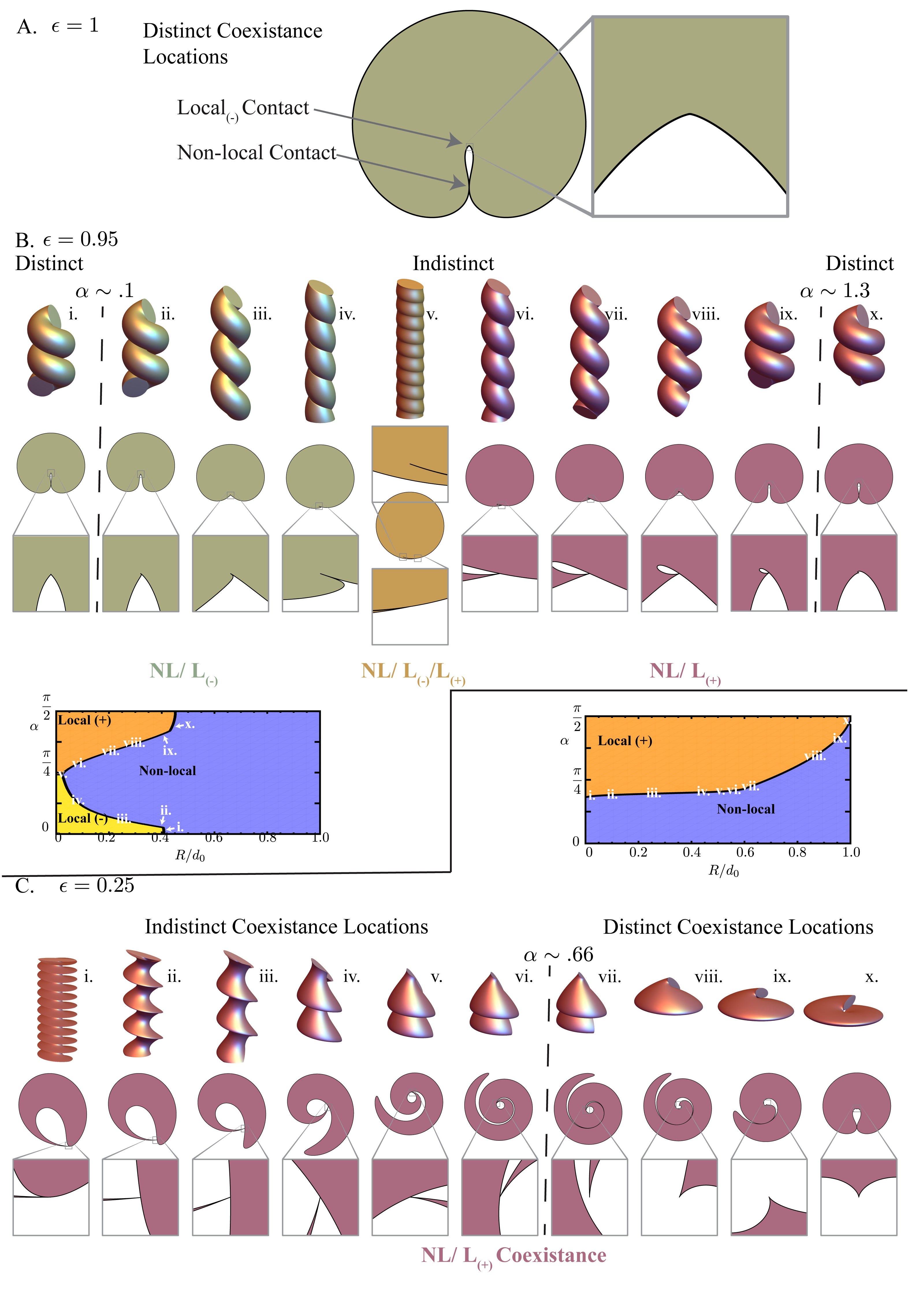}
    \caption{Coexistence between Local (L) and Non-local (NL) contact. (A.) Single point of isotropic ($\epsilon = 1$) L/NL coexistence; here, the local contact points ``inward'' (hence classified as local(-)) and the contacts occur at {\it distinct} locations along the tube surface.
    (B.) States along the L/NL coexistence boundary for slightly anisotropic tubes ($\epsilon = 0.95$); the boundary pans both local(-) and local(+), as discussed in Section~\ref{sec:highEps}. Most states (i.e. intermediate $\alpha$)coexistent contact at a {\it indistinct} (i.e. the same) location along the tube surface exhibit; extreme values of $\alpha$ exhibit distinct contact.
    (C.) States along the L/NL coexistence boundary for highly anisotropic tubes ($\epsilon = 0.25$). States at smaller $\alpha$ exhibit indistinct contact while states at larger $\alpha$ exhibit distinct contact. 
    }
    \label{fig:coxistance}
\end{figure*}

As reported in \cite{Pieranski_2001}, the isotropic filament contact regimes meet a single point, $R \simeq 0.431 d_0$, $P \simeq 1.083 d_0$. This structure, shown in Fig~\ref{fig:iso_p}A.iv simultaneously exhibits local and non-local. Perhaps intuitively, these two modes occur at {\it distinct} locations along the tube surface (i.e distinct $\psi$ coordinates); the local contact (or curvature constraint) occurs inside the helical core (i.e in the direction of curvature) while the non-local contact occurs between successive turns (i.e roughly in the vertical direction). Several anisotropic filaments with local/non-local coexistence were presented in the main text (Fig.~\ref{fig:e095_p}C.ii, Fig.~\ref{fig:e095_p}C.iv, Fig.~\ref{fig:e08175_p}C.ii, Fig.~\ref{fig:e08175_p}C.v, and Fig.~\ref{fig:e025_p}C.iii), however all of these structures appear to exhibit simultaneous contact at a single point in the croiss-section; the two sites of contact are {\it indistinct}. This curious difference occurs due to the specific pathway shown in Fig.~\ref{fig:e095_p}C - \ref{fig:e025_p}C, in particular that the coexistent structure is shown at a relatively low value for $R$. In Fig.~\ref{fig:coxistance}, we show additional structures along the local/non-local coexistence line. For both $\epsilon = 0.95$ and $\epsilon = 0.25$, we find that the boundary exhibits regions of distinct and indistinct coexistence, at large and small $R$, respectively; transitions between these motifs can be determined by inflections in the coexistence boundaries shown in Fig.~\ref{fig:e095_p}B and \ref{fig:e025_p}B. It seems then that this indistinct coexistence is then a consequence (or in fact, a cause) of the non-local contact regime's ability to persist to smaller helical radii; this structural motif is an inherent feature of cross-sectional anisotropy. 

\subsection{Nested packings at low $R_C$ and high $\epsilon$?}
Highly anistropic structures follow a straightforward progression, tape $\rightarrow$ scroll $\rightarrow$ screw as $R_C$ increases, as shown in e.g. Fig.~\ref{fig:phi_vs_R}A,bottom for $\epsilon = 0.25$. Curiously, when $\epsilon$ is large, we observe a second state of scroll-like packing, which is apparently more dense than the corresponding tape-like packing. This phenomena is observed under the entire $\epsilon$ scale where the Local(-) branch constrains the contact manifold, $1 > \epsilon \gtrsim 0.8175$, and corresponds for a preference for non-local contact over local(-) contact; we therefore dub this transition as the ``Cusp Escape'' regime.  
\begin{figure*}
    \centering
    \includegraphics[width=5in]{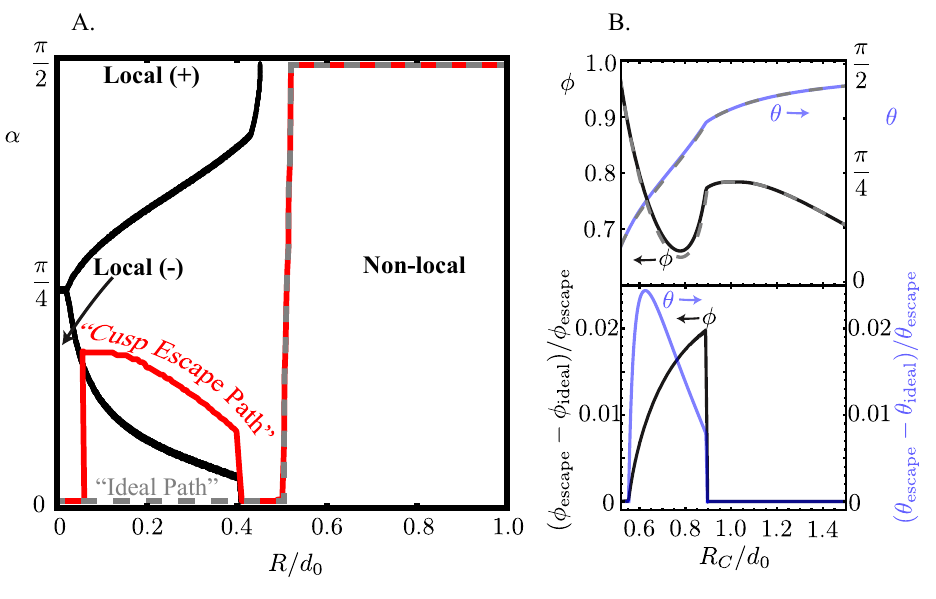}
    \caption{Optimal packing at high $\epsilon$ yields two nested regimes. (A.) $\epsilon = 0.95$ contact mode phase diagram (black, see Fig.~\ref{fig:e095_p}B), optimal packing (red, corresponding to Fig.~\ref{fig:phi_vs_R}A), and the ``Ideal'' low $\epsilon$ tape $\rightarrow$ scroll $\rightarrow$ screw progression (gray, dashed). The preferred packing exhibits intermediate tilt to ``escape'' the local(-) regime. (B. top) $\phi$ (black) and $\theta$ (blue) along the optimal branch, as well as along the ``Ideal'' branch (gray, dashed). (B. bottom) Normalized deviation of the quantities ($q$), $\phi$ and $\theta$.
    }
    \label{fig:cusp_escape}
\end{figure*}

Based on Fig.~\ref{fig:cusp_escape}B, it is obvious that this preference for intermediate tilt at low $R_C$ does not significantly influence the packing; for $\epsilon = 0.95$, the helical angle ($\theta$) and the packing density ($\phi$) only deviate by at most $\simeq 2 \%$; this largely is a consequence of the weak anisotropy. While this secondary tilt motif persists down to $\epsilon \simeq 0.8175$ (where tilt should have a larger consequence), the local(-) regime decrease in size with decreasing $\epsilon$ (see Fig.~\ref{fig:e08175_p}); as a result, the magnitude of $\alpha$ (and therefore deviation from ``Ideal'' packing) needed for ``Cusp Escape'' also decreases (compare, e.g. $\epsilon = 0.95$ and $\epsilon = 0.85$ in Fig.~\ref{fig:phi_vs_R}A. As such, we consider these low $R_C$/high $\epsilon$ nested states to be a relatively inconsequential deviation from the more intuitive tape $\rightarrow$ scroll $\rightarrow$ screw progression with increasing capillary size.

\section{\label{sec:App_phi_vs_RC}Capillary Contact Maximizes Density}

In the present work, we specifically report capillary density of dense (i.e. self-contacting) helices within their tightest capillary. However, our approach is in fact far more general; the capillary density, given in expression  Eqn.~\ref{eqn:phi} can be applied to helices of arbitrary configuration (in particular, ``expanded helices'' with $P > P_{\rm min}$) within arbitrary capillaries (in particular, with $R_C \geq {\rm max}_\psi |{\bf X}_\perp (\psi)|$).

While this larger ensemble of states may certainly be of interest in some physical phenomena or systems, our current focus is on states that \emph{maximize density}. 

It can be shown analytically that

\begin{equation}
   \frac{\partial \phi}{\partial P} \leq 0 
\end{equation}

for $P \geq P_{\rm min}$ (independent of the values of $R$, $\alpha$, $\epsilon$), so in surveying the densest configurations, it is appropriate to only consider packings which simultaneously exhibit self- and capillary-contact.

\section*{References}
\bibliographystyle{iopart-num}
\bibliography{refs}

\end{document}